\documentclass[prl,twocolumn,10pt,showpacs,superscriptaddress,longbibliography,footnoteinbib]{revtex4-1} 
\usepackage{amsfonts,amsmath,amssymb,graphicx}

\usepackage{color,times}

\definecolor{darkblue}{rgb}{0, 0, 0.8}
\usepackage[colorlinks=true, breaklinks=true, linkcolor=darkblue, citecolor=darkblue, urlcolor=darkblue]{hyperref}
\newcommand{\ket}[1]{|#1\rangle}

\begin{document}

\title{Measurement of the Angular Dependence of the Dipole-Dipole Interaction \\ Between Two Individual Rydberg Atoms at a F\"orster Resonance}

\author{Sylvain Ravets}
\author{Henning Labuhn}
\author{Daniel Barredo}
\author{Thierry Lahaye}
\author{Antoine Browaeys}

\affiliation{Laboratoire Charles Fabry, UMR 8501, Institut d'Optique, CNRS, Univ Paris Sud 11,\\
2 avenue Augustin Fresnel, 91127 Palaiseau cedex, France }


\begin{abstract}

We measure the angular dependence of the resonant dipole-dipole interaction between two individual Rydberg atoms with controlled relative positions. By applying a combination of static electric and magnetic fields on the atoms, we demonstrate the possibility to isolate a single interaction channel at a F\"orster resonance, that shows a well-defined angular dependence. We first identify spectroscopically the F\"orster resonance of choice and we then perform a direct measurement of the interaction strength between the two atoms as a function of the angle between the internuclear axis and the quantization axis. Our results show good agreement with the angular dependence $\propto(1-3\cos^2\theta)$ expected for this resonance. Understanding in detail the angular dependence of resonant interactions is important in view of using F\"orster resonances for quantum state engineering with Rydberg atoms.

\end{abstract}

\maketitle 

Among the systems currently considered for quantum engineering with applications to quantum simulation \cite{Georgescu2014} and quantum information \cite{Jones2012}, ensembles of individual Rydberg atoms are promising \cite{Jaksch2000,Lukin2001,Weimer2010}, since they provide large interactions \cite{Gallagher1994} and are scalable \cite{Saffman2010}. For that purpose, one needs to control pairwise interactions in the system. Significant achievements using small numbers of atoms have been obtained based on the phenomenon of Rydberg blockade \cite{Urban2009, Gaetan2009, Wilk2010, Isenhower2010,Comparat2010,Barredo2014,Hankin2014,Jau2015}. The energy transfer between atoms \cite{Anderson1998,Mourachko1998} observed in the presence of resonant interactions also leads to interesting many-body dynamics governed by spin-exchange Hamiltonians \cite{Gunter2013,Barredo2014b}. In those examples, one has to pay attention to the angular dependence of the interaction \cite{Reinhard2007,Walker2008,Marcassa2014,Vermersch2015}.

The anisotropy of the dipole-dipole interaction plays a crucial role in solid-state nuclear magnetic resonance \cite{Slichter1990}, and in systems such as cold polar molecules \cite{Ni2009,Hazzard2014} or dipolar quantum gases \cite{Lahaye2009}. Although experimental evidence of the anisotropy of the resonant dipole-dipole interaction between Rydberg atoms has been observed in confined geometries \cite{Carroll2004}, to date, a direct measurement of this angular dependence has not been obtained due to the complexity introduced by multiple interaction channels and inhomogeneities in disordered ensembles \cite{Reinhard2007,Cabral2011}. Here we use a combination of electric and magnetic fields to isolate a single interaction channel, by tuning two atoms separated by a controlled distance $R$ to a F\"orster resonance. We measure the evolution of the system~\cite{Ravets2014} to extract the interaction strength as a function of the angle $\theta$ between the interatomic axis and the quantization axis $z$. Our results further extend the possibilities offered by Rydberg atoms for quantum state engineering.

The electric dipole-dipole interaction between two atoms ($k=1,2$) is described by the operator
\begin{eqnarray}
\label{Eq:Vdd}
\hat{V}_{\rm dd} &=& \frac{1}{4 \pi \varepsilon_0 R^3 } \left[ \mathcal{A}_1 (\theta) \left( \hat{d}_{1+} \hat{d}_{2-} + \hat{d}_{1-} \hat{d}_{2+} + 2\hat{d}_{1z} \hat{d}_{2z} \right) \right. \nonumber \\
&& + \mathcal{A}_2 (\theta) \left( \hat{d}_{1+}\hat{d}_{2z} - \hat{d}_{1-}\hat{d}_{2z} +\hat{d}_{1z}\hat{d}_{2+} - \hat{d}_{1z}\hat{d}_{2-} \right) \nonumber \\
&& - \left. \mathcal{A}_3 (\theta) \left( \hat{d}_{1+}\hat{d}_{2+} + \hat{d}_{1-}\hat{d}_{2-} \right) \right] \ , 
\end{eqnarray}
where $\hat{d}_{k,\pm} = \mp( \hat{d}_{k,x} \pm i \hat{d}_{k,y} ) / \sqrt{2}$ are the components of the dipole operator in the spherical basis. The operator $\hat{V}_{\rm dd}$ in Eq.~\eqref{Eq:Vdd} contains terms with angular prefactors $\mathcal{A}_1(\theta) = (1-3 \cos ^2 \theta)/2$, $\mathcal{A}_2 (\theta) = 3 \sin \theta \cos \theta / \sqrt{2}$ and $\mathcal{A}_3 (\theta) = 3 \sin ^2 \theta / 2$, that couple pair states where the total magnetic quantum number $M=m_1+m_2$ changes by $\Delta M =0$, $\Delta M = \pm 1$ and $\Delta M = \pm 2$, respectively. In the absence of external fields, two atoms prepared in the same state generally interact in the van der Waals regime, where several states contribute to second order to the interaction \cite{Saffman2010}. Resonant dipole-dipole interactions occur when two pair states coupled by $\hat{V}_{\rm dd}$ are degenerate, giving rise to stronger interaction energies $E_{\rm dd} \propto 1/R^{3}$ \cite{Gallagher1994,Saffman2010}.

Resonant interactions between Rydberg atoms can be observed by applying an electric field, to reach a F\"orster resonance \cite{Walker2005, Anderson2002, Mudrich2005, Vogt2007, Ditzhuijzen2008, Ryabtsev2010, Nipper2012, Nipper2012b}. Here we study the resonance
\begin{equation}
\label{Eq:channel}
\begin{aligned}
59D_{3/2} + 59D_{3/2} \longleftrightarrow 61P_{1/2} + 57F_{5/2} \
\end{aligned}
\end{equation}
for two Rubidium atoms prepared in $\ket{d} = \ket{59D_{3/2},m_j=3/2}$ \cite{Ravets2014}. When $\theta=0$, the angular prefactors $\mathcal{A}_2$ and $\mathcal{A}_3$ vanish, and we only need to consider dipolar couplings with $\Delta M = 0$. In this case, Eq.~\eqref{Eq:channel} describes the transition of one atom to the magnetic substate $\ket{p}=\ket{61P_{1/2},m_j=1/2}$ while the other atom evolves to the magnetic substate $\ket{f_1}=\ket{57F_{5/2},m_j=5/2}$. We define the symmetric state $\ket{pf_1}_{\rm s} = \left( \ket{pf_1}+\ket{f_1p} \right) / \sqrt{2}$. In a zero electric field, the energy splitting between $\ket{dd}$ and $\ket{pf_1}_{\rm s}$ is only $h\times8.5$~MHz, and vanishes for a small electric field ($F_{\rm res} \simeq 33~{\rm mV/cm}$) due to the differential Stark effect between the two states~\footnote{The electric field required to reach the resonance is weak enough to be in a regime of induced dipoles, by opposition to rigid dipoles.}. At resonance, the eigenstates of the system are the combinations $\ket{\pm_1} = \left( \ket{dd} \pm \ket{pf_1}_{\rm s} \right) / \sqrt{2}$, that have energies $E_{\pm}=\pm \sqrt{2} C_3 / R^3$. The coefficient $C_3$, defined such that the matrix element of $\hat{V}_{\rm dd}$ between $\ket{dd}$ and $\ket{pf_1}$ for $\theta=0$ is equal to $C_3/R^3$, was measured previously in \cite{Ravets2014} to be $C_3 / h = 2.39 \pm 0.03~{\rm GHz.\mu m^3}$. 

\begin{figure}[t]
\centering
\includegraphics[width=86mm]{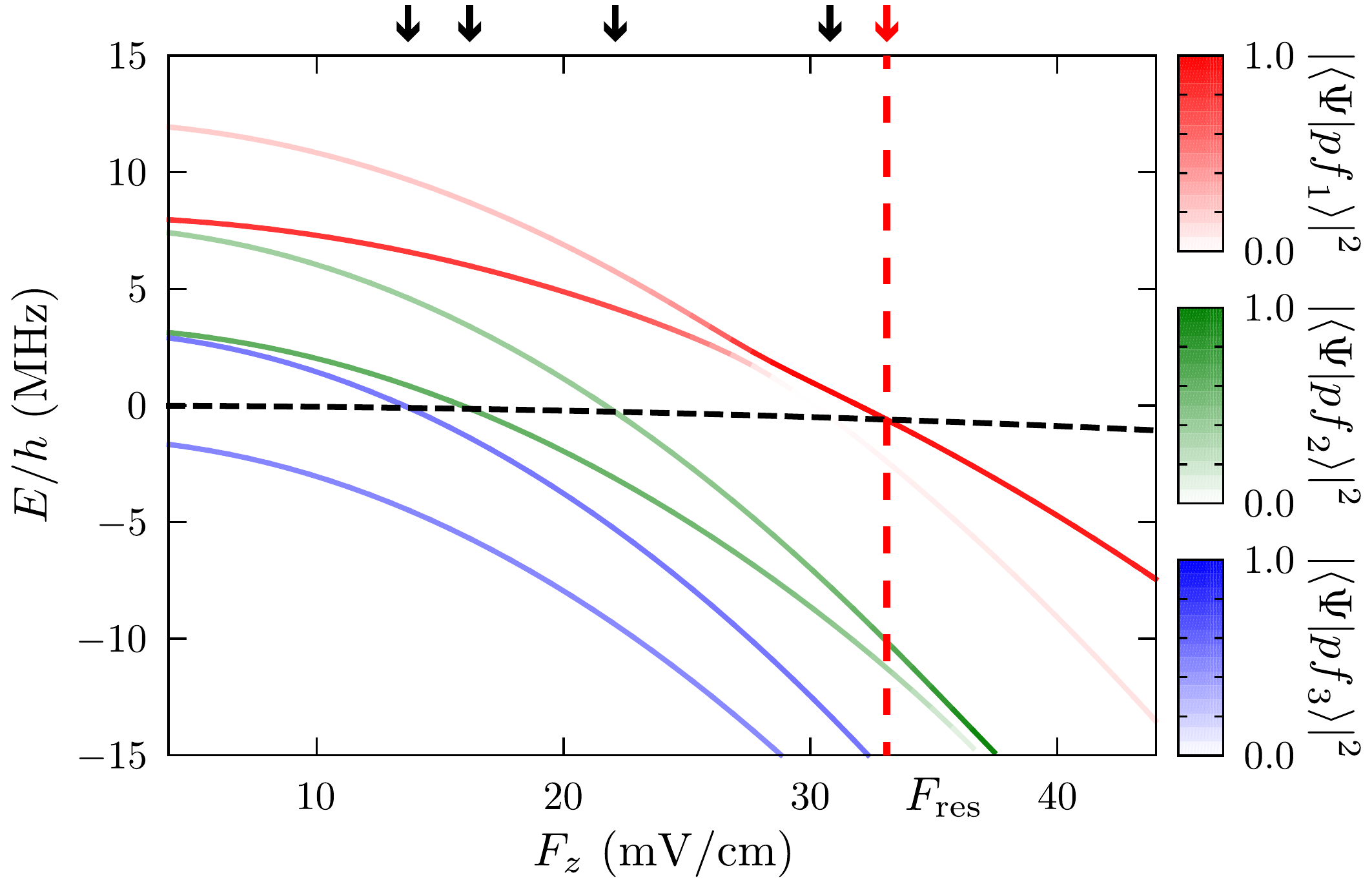}
\caption{(color online) Stark map of the relevant pair states in the presence of a $3.3~\rm{G}$ magnetic field. The colored lines plot the eigenenergies of the two-atom system as a function of the electric field. The color scale represents the overlap between the calculated eigenstates $\ket{\Psi}$ and the states$\ket{pf_1}_{\rm s}$, $\ket{pf_2}_{\rm s}$, $\ket{pf_3}_{\rm s}$. The black dashed line shows the energy of $\ket{dd}$. The solid arrows show the positions of the different resonances. The red vertical dashed line indicates the position of the resonance detailed in this paper.}
\label{fig:Resonance}
\end{figure}

Here, we focus on the angular dependence of this interaction. When $\theta \neq 0$, $\hat{V}_{\rm dd}$ couples $\ket{dd}$ to different magnetic substates $\ket{pf_i}_{\rm s}$, as detailed later (see Fig.~\ref{fig:Resonance}). Since the states $\ket{pf_i}_{\rm s}$ have different polarizabilities, it is possible, by a proper choice of parameters, to isolate one interaction channel~\cite{Nipper2012,Nipper2012b} showing a well-defined angular dependence. This is what we demonstrate here for the resonance between $\ket{pf_1}_{\rm s}$ and $\ket{dd}$, with full control over the geometry of the system, which allows us to measure an interaction strength varying as $\mathcal{A}_1(\theta)$.

Our experiment starts by loading two $^{87} {\rm Rb}$ atoms from a magneto-optical trap into two microscopic dipole traps, created by focusing a laser beam of wavelength $852~{\rm nm}$ using a high-numerical aperture lens \cite{Beguin2013}. We control $R$ and $\theta$ using a spatial light modulator to create arbitrary trap patterns \cite{Nogrette2014}. Using a set of electrodes, we control the electric field $F_z$ at the position of the atoms. We apply a $3.3~{\rm G}$ magnetic field along $z$, and we optically pump the atoms in $\ket{g} = \ket{5S_{1/2},F=2,m_F=2}$. For the Rydberg excitation from $\ket{g}$ to $\ket{d} = \ket{59D_{3/2},m_j=3/2}$, we perform a two-photon transition using a $\pi$-polarized laser beam of wavelength $795~{\rm nm}$ and a $\sigma ^ +$-polarized laser beam of wavelength $474~{\rm nm}$. During that time, the microtraps are switched off. At the end of the sequence, we turn on the trapping light again and shine resonant light of wavelength $780~{\rm nm}$ on the atoms. We read out the state of the system by detecting the fluorescence of ground-state atoms (atoms that are in a Rydberg state at the end of the excitation sequence are not recaptured in the microtraps). We repeat the sequence $\simeq 100$ times to reconstruct the two-atom state populations $P_{gg}$, $P_{gr}$, $P_{rg}$ and $P_{rr}$, where $r$ stands for any Rydberg state and $g$ stands for the ground state.

\begin{figure}[t]
\centering
\includegraphics[width=86mm]{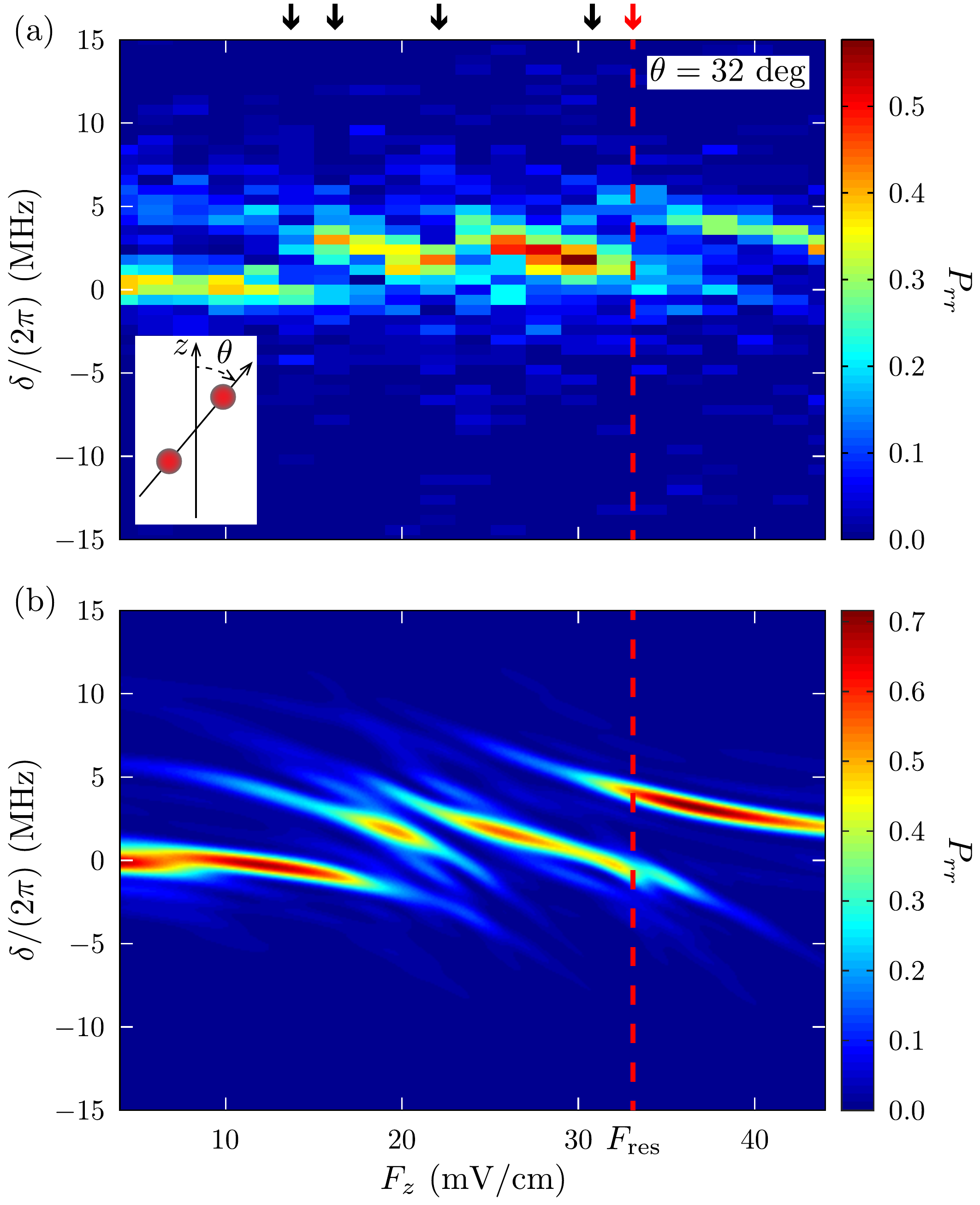}
\caption{(color online) Rydberg excitation spectrum for two atoms with $R=9.1~{\rm \mu m}$ and $\theta = 32 ^{\circ}$. The double excitation probability $P_{rr}$ is represented as a function of the detuning $\delta$ and the electric field $F_z$. The red vertical dashed line shows the measured position of the F\"orster resonance detailed in this paper. The bold arrows show the calculated positions of the different resonances. (a)~Experimental spectrum. (b)~Theoretical spectrum obtained by solving the Schr\"odinger equation for the system (see text).}
\label{fig:Spectroscopy}
\end{figure}

In a first experiment, we identify the various resonances by measuring Rydberg excitation spectra as a function of the electric field for two atoms with $R=9.1~{\rm \mu m}$ and $\theta = 32^\circ$. For electric fields that vary between $F_z = 4~{\rm mV/cm}$ and $F_z = 44~{\rm mV/cm}$, we shine on the atoms an excitation laser pulse of Rabi frequency $\Omega / 2 \pi \simeq 0.76~{\rm MHz}$ and duration $\tau=1~\rm{\mu s}$. We record the probability $P_{rr}$ to excite both atoms to a Rydberg state as we scan the frequency detuning $\delta$ of the excitation laser with respect to the Rydberg line measured in a zero electric field. Figure~\ref{fig:Spectroscopy}(a) shows the result of this measurement. For electric fields smaller than a few mV/cm, we observe a single excitation line showing the transfer of population from $\ket{gg}$ to $\ket{dd}$ via the pair states $\ket{gd}$ and $\ket{dg}$. At resonance between $\ket{dd}$ and $\ket{pf_i}_{\rm s}$, the state $\ket{gg}$ is partially coupled to the eigenstates $\ket{\pm _i}=(\ket{dd} \pm \ket{pf_i}_{\rm s})$, and we observe the presence of two excitation lines. Because $\ket{pf_i}_{\rm s}$ have different polarizabilities, resonances between $\ket{dd}$ and $\ket{pf_i}_{\rm s}$ occur at distinct $F_z$, and we observe different avoided crossings between $\ket{dd}$ and $\ket{pf_i}_{\rm s}$. The measured spectrum is in good agreement with the calculated spectrum shown in Fig.~\ref{fig:Spectroscopy}(b), obtained by simulating the experimental sequence for the system, as detailed later in this paper (see Fig.~\ref{fig:Resonance}). For $\theta = 0$, the same measurement (not shown here) gives a single avoided crossing between $\ket{dd}$ and $\ket{pf_1}_{\rm s}$~\cite{Ravets2014}. The observation of multiple resonances when $\theta \neq 0$ is a first indication of the anisotropy of the interaction.

We now focus on the resonance at $F_z \simeq 33~{\rm mV/cm}$ (red dashed line on Fig.~\ref{fig:Spectroscopy}) and study its angular dependence. For a given angle $\theta$, we obtain the interaction strength between the two atoms by measuring the time evolution of the system in the presence of resonant interactions \cite{Ravets2014}. In this experiment, $R=9.1~{\rm \mu m}$, $\Omega / 2 \pi \simeq 5~{\rm MHz}$ and $\theta$ varies between $-90^{\circ}$ and $90^{\circ}$. We first place the system out of resonance ($F_z = 64~{\rm mV/cm}$), where the atoms show weak van der Waals interactions $\leq 1~{\rm MHz}$. We apply an optical $\pi$-pulse to prepare the two-atom system in the state $\ket{dd}$. We then switch on abruptly strong resonant interactions by pulsing the electric field to $33~{\rm mV/cm}$ with rise and fall times $\leq 10~{\rm ns}$, thus inducing back and forth oscillations between $\ket{dd}$ and $\ket{pf_1}_{\rm s}$. After an interaction time $t$, the probability for both atoms to be in the state $\ket{d}$ reads
\begin{equation}
\begin{aligned}
P_{dd}(t) = \cos ^2 \left(\frac{\sqrt{2} C_3 \mathcal{A}_1(\theta)}{\hbar R^3} t \right) \ ,
\end{aligned}
\end{equation}
where we neglect the influence of other interaction channels. At the end of the sequence, we freeze the dynamics by switching the electric field back to $F_z= 64~{\rm mV/cm}$. To read out the state of the system, we shine a second optical $\pi$-pulse on the atoms, that couples $\ket{dd}$ back to $\ket{gg}$ while leaving $\ket{pf_1}_{\rm s}$ unchanged. We measure the probability $P_{gg}$ to detect both atoms in the ground state. Assuming perfect excitation and detection, $P_{gg}$ coincides with $P_{dd}$.

\begin{figure}[t]
\centering
\includegraphics[width=86mm]{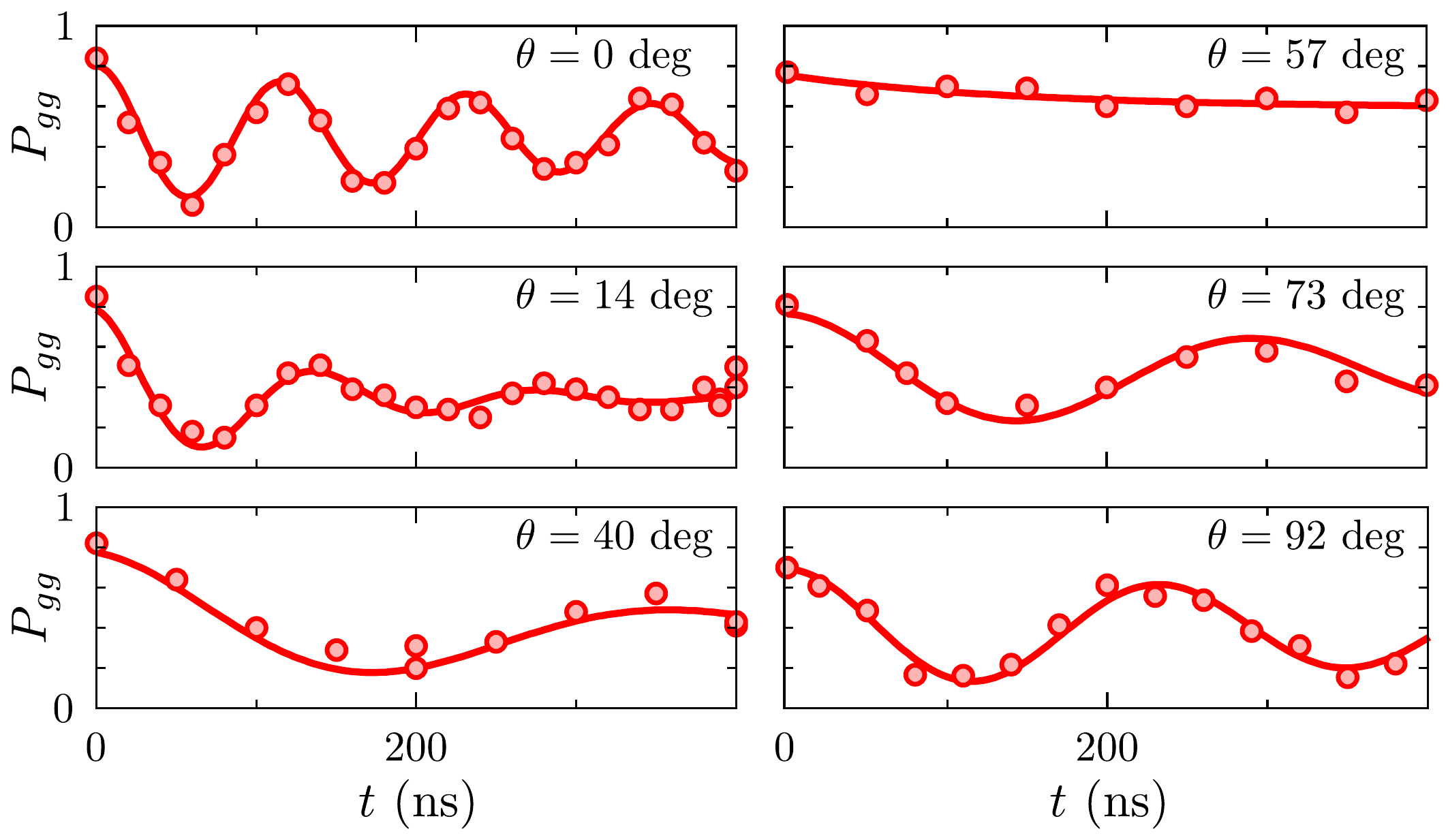}
\caption{(color online) Coherent oscillations at resonance for angles varying between $\theta = 0$ and $\theta = 92^{\circ}$, where $R=9.1~\rm{\mu m}$. The typical error on $P_{gg}$ is $\pm 0.05$. Lines are fits to the data by damped sines (see text).}
\label{fig:Oscillation}
\end{figure}

Figure~\ref{fig:Oscillation} shows the coherent oscillations observed for different angles $\theta$. The anisotropy of the interaction is evident on the data. For $\theta = 0$ we observe contrasted oscillations with frequency $\nu \simeq 8.6 \pm 0.4~{\rm MHz}$, compatible with the value of $C_3$ measured using the same experimental sequence and varying~$R$~\cite{Ravets2014}. As we increase~$\theta$, we initially observe a decrease of~$\nu$. For $\theta \simeq 57^{\circ}$, the observed dynamics are essentially described by an exponential damping. This is compatible with the expected behavior close to the ``magic angle'' $(\theta_{\rm m} \simeq 54.7^\circ)$ where the strength of resonant interactions vanishes ($\mathcal{A}_1 (\theta_{\rm m}) = 0$), and where one thus expects $P_{dd}$ to stay constant. Further increasing $\theta$ leads again to the observation of contrasted oscillations with increasing frequencies. The observed damping originates mainly from dephasing, due to the finite temperature of the atoms and fluctuations in the voltages applied to the electrodes leading, respectively, to shot-to-shot fluctuations of $R$ and of the detuning between $\ket{dd}$ and $\ket{pf_1}_{\rm s}$. The observation of mainly one frequency in the oscillations confirms that the interaction channel of choice is well-isolated.

\begin{figure}[t]
\centering
\includegraphics[width=86mm]{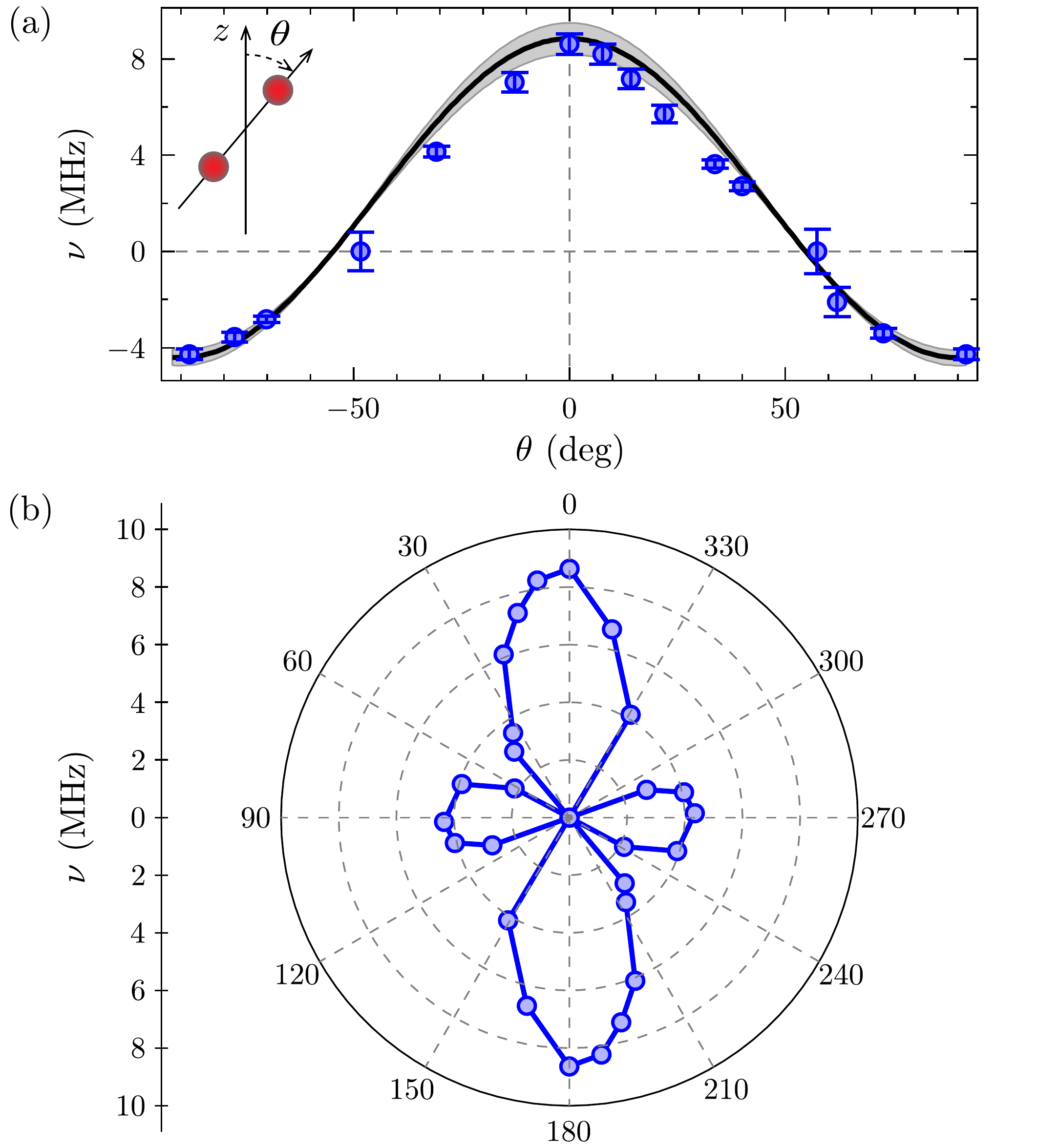}
\caption{(color online) Angular dependence of the interaction. (a)~Graph of $\nu$ as a function of $\theta$. The solid line plots $- C_3 \mathcal{A}_1 (\theta)/(h R^3)$. Error bars show statistical errors in the determination of $\nu$. The shaded area corresponds to a systematic 5\% error on the calibration of $R$. (b)~Representation in polar coordinates. By symmetry, the points at angles at $\theta + 180^{\circ}$ are taken identical to the points at $\theta$.}
\label{fig:AngularDependence}
\end{figure}

We obtain the interaction strength as a function of $\theta$ by fitting the measured coherent oscillations with damped sines of frequency $\nu$. Figure~\ref{fig:AngularDependence}(a) shows a plot of $\nu$ as a function of $\theta$. Our measurement procedure does not give access to the sign of the interaction, and we thus infer it from the expected functional form. The error bars represent statistical errors from the fit. For $\theta \simeq 57^{\circ}$, close to the magic angle, we do not observe any significant oscillation within the limits of experimental accuracy. In this case, we assign the value 0~MHz to the frequency, with an error bar representing the upper frequency one could possibly infer from the measurement and equal to $1/(2t_{\rm max})$, where $t_{\rm max} = 700~{\rm ns}$ is our largest interaction time. The data show good agreement with the expected angular dependence $\propto \mathcal{A}_1 (\theta)$ plotted as a solid line in Fig.~\ref{fig:AngularDependence}(a), where we used the coefficient $C_3 $ measured in \cite{Ravets2014}. The polar representation of the fitted frequencies as a function of $\theta$ exhibits the characteristic shape of the function $(1-3 \cos ^2 \theta)$, as can be observed in Fig.~\ref{fig:AngularDependence}(b).

In the remaining of this paper, we analyze theoretically our results. We first consider the situation where no magnetic field is applied on the atoms. Coming back to Eq.~\eqref{Eq:Vdd}, terms with the angular prefactor $\mathcal{A}_1(\theta)$ couple $\ket{dd}$ and $\ket{pf_1}_{\rm s}$ ($\Delta M = 0$). Terms with the angular prefactor $\mathcal{A}_2 (\theta)$ couple $\ket{dd}$ and $\ket{pf_2}_{\rm s}$, with $\ket{f_2} = \ket{57F_{5/2},m_j=3/2}$ ($\Delta M = -1$). Finally, terms with the angular prefactor $\mathcal{A}_3 (\theta)$ couple $\ket{dd}$ and $\ket{pf_3}_{\rm s}$, with $\ket{f_3} = \ket{57F_{5/2},m_j=1/2}$ ($\Delta M = -2$). In the absence of magnetic field, we thus expect three resonances between $\ket{dd}$ and the states $\ket{pf_1}$, $\ket{pf_2}$ and $\ket{pf_3}$.

We now take into account the effect of the $3.3~{\rm G}$ magnetic field applied on the atoms during our experiments, leading to energy shifts and to mixing of the different states \cite{Nipper2012, Nipper2012b}. Figure~\ref{fig:Resonance} shows the calculated Stark map for the relevant pair states, where the dipole-dipole interaction has not been included. It was obtained by calculating the (one-atom) level shifts of the states $\ket{p}$, $\ket{d}$, and $\ket{f_i}$ from the diagonalization of the total Hamiltonian, including the Stark and Zeeman effects~\footnote{For the Stark map calculation of the states \unexpanded{$\ket{f_i}$}, we found it necessary to include a large number of states in the basis to achieve convergence. Note that, due to the proximity of the hydrogenic manifold, the $B=3.3~{\rm G}$ magnetic field cannot be taken into account perturbatively in the calculation. Here, we thus used a basis containing 756~states, with $3 \leq L \leq 56$, $-3 \leq m_L \leq 3$ and $-1/2 \leq m_S \leq 1/2$.}. The electric dipole matrix elements were obtained numerically using quantum defect theory \cite{Li2003,Han2006,Reinhard2007}. We observe that the magnetic field shifts the pair states energies due to the Zeeman effect, and mixes $\ket{pf_i}_{\rm s}$ to $\ket{pf_{i+3}}_{\rm s}$ ($i=1,2,3$), where $\ket{f_4} = \ket{57F_{7/2},m_j=5/2}$, $\ket{f_5} = \ket{57F_{7/2},m_j=3/2}$ and $\ket{f_6} = \ket{57F_{7/2},m_j=1/2}$ (see solid lines paired by colors in Fig.~\ref{fig:Resonance}). Because all pair states have different polarizabilities, the system exhibits a discrete set of F\"orster resonances (see bold arrows), at the crossings between the solid colored lines and the black dashed line. The color bar represents the degree of admixture of the different states and shows that the presence of a magnetic field increases the number of expected resonances.

Based on this Stark map, we finally simulated numerically the experimental spectrum of the two-atom system of Fig.~\ref{fig:Spectroscopy}(a). For that purpose, we first truncated the two-atom basis, retaining only the two-atom states differing in energy by no more than 40~MHz from $\ket{dd}$, and then calculated the restriction of $\hat{V}_{\rm dd}$ to this truncated basis. The result of the calculation, shown in Fig.~\ref{fig:Spectroscopy}(b), is in good agreement with the experiment regarding the positions and the strengths of the different resonances. For our experimental parameters, we note that we resolve three avoided crossings. We focused in this paper on the crossing between the states $\ket{dd}$ and $\ket{pf_1}_{\rm s}$ occurring at $F_z \simeq 33 ~ \rm{mV/cm}$. In this case, the states $\ket{pf_2}_{\rm s}$ and $\ket{pf_3}_{\rm s}$ are off-resonant by more than 10~{\rm MHz}, which ensures that this resonance is well isolated from other interaction channels, as observed on the interaction dynamics. On the contrary, in the absence of magnetic field, the energy splitting between the different states ($\leq 10~{\rm MHz}$) would not be sufficient to isolate the different channels. Using a finite magnetic field was thus necessary to isolate a single resonance. By reproducing the same procedure with different interaction channels, one could in principle engineer other angular dependencies $\propto \mathcal{A}_2(\theta)$ or $\propto \mathcal{A}_3(\theta)$.

The control over the anisotropy of the dipole-dipole interaction demonstrated in this work will be useful for the implementation of quantum information protocols \cite{Muller2009} or for the quantum simulation of many-body spin systems in large 2D arrays of dipole traps \cite{Nogrette2014}. It is also a motivation to revisit, in the presence of anisotropic interactions, effects discussed previously such as the prediction of the breakdown of the Rydberg blockade in the presence of nearly resonant interactions \cite{Pohl2009}, the existence of few-body F\"orster resonances \cite{Gurian2012} or the diffusion of spin excitations in complex ensembles of spins with long-range interactions \cite{Cote2006,Wuster2010}.

\begin{acknowledgments}
We thank T.~Pfau and P.~Cheinet for insightful discussions. We acknowledge financial support by the EU [ERC Stg Grant ARENA, Project HAIRS, H2020 FET Proactive project RySQ (grant N. 640378), Marie-Curie Program ITN COHERENCE FP7-PEOPLE-2010-ITN-265031 (H. L.)], and by the PALM Labex [project QUANTICA].
\end{acknowledgments}


\begin{thebibliography}{49}%
\makeatletter
\providecommand \@ifxundefined [1]{%
 \@ifx{#1\undefined}
}%
\providecommand \@ifnum [1]{%
 \ifnum #1\expandafter \@firstoftwo
 \else \expandafter \@secondoftwo
 \fi
}%
\providecommand \@ifx [1]{%
 \ifx #1\expandafter \@firstoftwo
 \else \expandafter \@secondoftwo
 \fi
}%
\providecommand \natexlab [1]{#1}%
\providecommand \enquote  [1]{#1}%
\providecommand \bibnamefont  [1]{#1}%
\providecommand \bibfnamefont [1]{#1}%
\providecommand \citenamefont [1]{#1}%
\providecommand \href@noop [0]{\@secondoftwo}%
\providecommand \href [0]{\begingroup \@sanitize@url \@href}%
\providecommand \@href[1]{\@@startlink{#1}\@@href}%
\providecommand \@@href[1]{\endgroup#1\@@endlink}%
\providecommand \@sanitize@url [0]{\catcode `\\12\catcode `\$12\catcode
  `\&12\catcode `\#12\catcode `\^12\catcode `\_12\catcode `\%12\relax}%
\providecommand \@@startlink[1]{}%
\providecommand \@@endlink[0]{}%
\providecommand \url  [0]{\begingroup\@sanitize@url \@url }%
\providecommand \@url [1]{\endgroup\@href {#1}{\urlprefix }}%
\providecommand \urlprefix  [0]{URL }%
\providecommand \Eprint [0]{\href }%
\providecommand \doibase [0]{http://dx.doi.org/}%
\providecommand \selectlanguage [0]{\@gobble}%
\providecommand \bibinfo  [0]{\@secondoftwo}%
\providecommand \bibfield  [0]{\@secondoftwo}%
\providecommand \translation [1]{[#1]}%
\providecommand \BibitemOpen [0]{}%
\providecommand \bibitemStop [0]{}%
\providecommand \bibitemNoStop [0]{.\EOS\space}%
\providecommand \EOS [0]{\spacefactor3000\relax}%
\providecommand \BibitemShut  [1]{\csname bibitem#1\endcsname}%
\let\auto@bib@innerbib\@empty
\bibitem [{\citenamefont {Georgescu}\ \emph {et~al.}(2014)\citenamefont
  {Georgescu}, \citenamefont {Ashhab},\ and\ \citenamefont
  {Nori}}]{Georgescu2014}%
  \BibitemOpen
  \bibfield  {author} {\bibinfo {author} {\bibfnamefont {I.~M.}\ \bibnamefont
  {Georgescu}}, \bibinfo {author} {\bibfnamefont {S.}~\bibnamefont {Ashhab}}, \
  and\ \bibinfo {author} {\bibfnamefont {F.}~\bibnamefont {Nori}},\ }\bibfield
  {title} {\enquote {\bibinfo {title} {{Q}uantum simulation},}\ }\href {\doibase 10.1103/RevModPhys.86.153} {\bibfield  {journal} {\bibinfo
  {journal} {Rev. Mod. Phys.}\ }\textbf {\bibinfo {volume} {86}},\ \bibinfo
  {pages} {153} (\bibinfo {year} {2014})}\BibitemShut {NoStop}%
\bibitem [{\citenamefont {Jones}\ and\ \citenamefont
  {Jaksch}(2012)}]{Jones2012}%
  \BibitemOpen
  \bibfield  {author} {\bibinfo {author} {\bibfnamefont {J.~A.}\ \bibnamefont
  {Jones}}\ and\ \bibinfo {author} {\bibfnamefont {D.}~\bibnamefont {Jaksch}},\
  }\href {http://dx.doi.org/10.1017/CBO9781139028509} {\emph {\bibinfo {title}
  {{Q}uantum {I}nformation, {C}omputation and {C}ommunication}}}\ (\bibinfo
  {publisher} {Cambridge University Press},\ \bibinfo {year}
  {2012})\BibitemShut {NoStop}%
\bibitem [{\citenamefont {Jaksch}\ \emph {et~al.}(2000)\citenamefont {Jaksch},
  \citenamefont {Cirac}, \citenamefont {Zoller}, \citenamefont {Rolston},
  \citenamefont {C\^ot\'e},\ and\ \citenamefont {Lukin}}]{Jaksch2000}%
  \BibitemOpen
  \bibfield  {author} {\bibinfo {author} {\bibfnamefont {D.}~\bibnamefont
  {Jaksch}}, \bibinfo {author} {\bibfnamefont {J.~I.}\ \bibnamefont {Cirac}},
  \bibinfo {author} {\bibfnamefont {P.}~\bibnamefont {Zoller}}, \bibinfo
  {author} {\bibfnamefont {S.~L.}\ \bibnamefont {Rolston}}, \bibinfo {author}
  {\bibfnamefont {R.}~\bibnamefont {C\^ot\'e}}, \ and\ \bibinfo {author}
  {\bibfnamefont {M.~D.}\ \bibnamefont {Lukin}},\ }\bibfield  {title} {\enquote
  {\bibinfo {title} {{F}ast {Q}uantum {G}ates for {N}eutral {A}toms},}\ }\href {\doibase 10.1103/PhysRevLett.85.2208} {\bibfield  {journal} {\bibinfo
  {journal} {Phys. Rev. Lett.}\ }\textbf {\bibinfo {volume} {85}},\ \bibinfo
  {pages} {2208} (\bibinfo {year} {2000})}\BibitemShut {NoStop}%
\bibitem [{\citenamefont {Lukin}\ \emph {et~al.}(2001)\citenamefont {Lukin},
  \citenamefont {Fleischhauer}, \citenamefont {Cote}, \citenamefont {Duan},
  \citenamefont {Jaksch}, \citenamefont {Cirac},\ and\ \citenamefont
  {Zoller}}]{Lukin2001}%
  \BibitemOpen
  \bibfield  {author} {\bibinfo {author} {\bibfnamefont {M.~D.}\ \bibnamefont
  {Lukin}}, \bibinfo {author} {\bibfnamefont {M.}~\bibnamefont {Fleischhauer}},
  \bibinfo {author} {\bibfnamefont {R.}~\bibnamefont {Cote}}, \bibinfo {author}
  {\bibfnamefont {L.~M.}\ \bibnamefont {Duan}}, \bibinfo {author}
  {\bibfnamefont {D.}~\bibnamefont {Jaksch}}, \bibinfo {author} {\bibfnamefont
  {J.~I.}\ \bibnamefont {Cirac}}, \ and\ \bibinfo {author} {\bibfnamefont
  {P.}~\bibnamefont {Zoller}},\ }\bibfield  {title} {\enquote {\bibinfo {title}
  {{D}ipole {B}lockade and {Q}uantum {I}nformation {P}rocessing in {M}esoscopic
  {A}tomic {E}nsembles},}\ }\href {\doibase 10.1103/PhysRevLett.87.037901}
  {\bibfield  {journal} {\bibinfo  {journal} {Phys. Rev. Lett.}\ }\textbf
  {\bibinfo {volume} {87}},\ \bibinfo {pages} {037901} (\bibinfo {year}
  {2001})}\BibitemShut {NoStop}%
\bibitem [{\citenamefont {Weimer}\ \emph {et~al.}(2010)\citenamefont {Weimer},
  \citenamefont {Muller}, \citenamefont {Lesanovsky}, \citenamefont {Zoller},\
  and\ \citenamefont {Buchler}}]{Weimer2010}%
  \BibitemOpen
  \bibfield  {author} {\bibinfo {author} {\bibfnamefont {H.}~\bibnamefont
  {Weimer}}, \bibinfo {author} {\bibfnamefont {M.}~\bibnamefont {Muller}},
  \bibinfo {author} {\bibfnamefont {I.}~\bibnamefont {Lesanovsky}}, \bibinfo
  {author} {\bibfnamefont {P.}~\bibnamefont {Zoller}}, \ and\ \bibinfo {author}
  {\bibfnamefont {H.~P.}\ \bibnamefont {Buchler}},\ }\bibfield  {title}
  {\enquote {\bibinfo {title} {{A} {R}ydberg quantum simulator},}\ }\href {http://dx.doi.org/10.1038/nphys1614} {\bibfield  {journal} {\bibinfo
  {journal} {Nat. Phys.}\ }\textbf {\bibinfo {volume} {6}},\ \bibinfo {pages}
  {382} (\bibinfo {year} {2010})}\BibitemShut {NoStop}%
\bibitem [{\citenamefont {Gallagher}(1994)}]{Gallagher1994}%
  \BibitemOpen
  \bibfield  {author} {\bibinfo {author} {\bibfnamefont {T.~F.}\ \bibnamefont
  {Gallagher}},\ }\href {http://dx.doi.org/10.1017/CBO9780511524530} {\emph
  {\bibinfo {title} {{R}ydberg {A}toms}}}\ (\bibinfo  {publisher} {Cambridge
  University Press},\ \bibinfo {year} {1994})\BibitemShut {NoStop}%
\bibitem [{\citenamefont {Saffman}\ \emph {et~al.}(2010)\citenamefont
  {Saffman}, \citenamefont {Walker},\ and\ \citenamefont
  {M\o{}lmer}}]{Saffman2010}%
  \BibitemOpen
  \bibfield  {author} {\bibinfo {author} {\bibfnamefont {M.}~\bibnamefont
  {Saffman}}, \bibinfo {author} {\bibfnamefont {T.~G.}\ \bibnamefont {Walker}},
  \ and\ \bibinfo {author} {\bibfnamefont {K.}~\bibnamefont {M\o{}lmer}},\
  }\bibfield  {title} {\enquote {\bibinfo {title} {{Q}uantum information with
  {R}ydberg atoms},}\ }\href {\doibase 10.1103/RevModPhys.82.2313} {\bibfield
  {journal} {\bibinfo  {journal} {Rev. Mod. Phys.}\ }\textbf {\bibinfo {volume}
  {82}},\ \bibinfo {pages} {2313} (\bibinfo {year} {2010})}\BibitemShut
  {NoStop}%
\bibitem [{\citenamefont {Urban}\ \emph {et~al.}(2009)\citenamefont {Urban},
  \citenamefont {Johnson}, \citenamefont {Henage}, \citenamefont {Isenhower},
  \citenamefont {Yavuz}, \citenamefont {Walker},\ and\ \citenamefont
  {Saffman}}]{Urban2009}%
  \BibitemOpen
  \bibfield  {author} {\bibinfo {author} {\bibfnamefont {E.}~\bibnamefont
  {Urban}}, \bibinfo {author} {\bibfnamefont {T.~A.}\ \bibnamefont {Johnson}},
  \bibinfo {author} {\bibfnamefont {T.}~\bibnamefont {Henage}}, \bibinfo
  {author} {\bibfnamefont {L.}~\bibnamefont {Isenhower}}, \bibinfo {author}
  {\bibfnamefont {D.~D.}\ \bibnamefont {Yavuz}}, \bibinfo {author}
  {\bibfnamefont {T.~G.}\ \bibnamefont {Walker}}, \ and\ \bibinfo {author}
  {\bibfnamefont {M.}~\bibnamefont {Saffman}},\ }\bibfield  {title} {\enquote
  {\bibinfo {title} {{O}bservation of {R}ydberg blockade between two atoms},}\
  }\href {http://dx.doi.org/10.1038/nphys1178} {\bibfield  {journal} {\bibinfo
  {journal} {Nat. Phys.}\ }\textbf {\bibinfo {volume} {5}},\ \bibinfo {pages}
  {110} (\bibinfo {year} {2009})}\BibitemShut {NoStop}%
\bibitem [{\citenamefont {Ga\"{e}tan}\ \emph {et~al.}(2009)\citenamefont
  {Ga\"{e}tan}, \citenamefont {Miroshnychenko}, \citenamefont {Wilk},
  \citenamefont {Chotia}, \citenamefont {Viteau}, \citenamefont {Comparat},
  \citenamefont {Pillet}, \citenamefont {Browaeys},\ and\ \citenamefont
  {Grangier}}]{Gaetan2009}%
  \BibitemOpen
  \bibfield  {author} {\bibinfo {author} {\bibfnamefont {A.}~\bibnamefont
  {Ga\"{e}tan}}, \bibinfo {author} {\bibfnamefont {Y.}~\bibnamefont
  {Miroshnychenko}}, \bibinfo {author} {\bibfnamefont {T.}~\bibnamefont
  {Wilk}}, \bibinfo {author} {\bibfnamefont {A.}~\bibnamefont {Chotia}},
  \bibinfo {author} {\bibfnamefont {M.}~\bibnamefont {Viteau}}, \bibinfo
  {author} {\bibfnamefont {D.}~\bibnamefont {Comparat}}, \bibinfo {author}
  {\bibfnamefont {P.}~\bibnamefont {Pillet}}, \bibinfo {author} {\bibfnamefont
  {A.}~\bibnamefont {Browaeys}}, \ and\ \bibinfo {author} {\bibfnamefont
  {P.}~\bibnamefont {Grangier}},\ }\bibfield  {title} {\enquote {\bibinfo
  {title} {{O}bservation of collective excitation of two individual atoms in
  the {R}ydberg blockade regime},}\ }\href {http://dx.doi.org/10.1038/nphys1183} {\bibfield  {journal} {\bibinfo
  {journal} {Nat. Phys.}\ }\textbf {\bibinfo {volume} {5}},\ \bibinfo {pages}
  {115} (\bibinfo {year} {2009})}\BibitemShut {NoStop}%
\bibitem [{\citenamefont {Wilk}\ \emph {et~al.}(2010)\citenamefont {Wilk},
  \citenamefont {Ga\"etan}, \citenamefont {Evellin}, \citenamefont {Wolters},
  \citenamefont {Miroshnychenko}, \citenamefont {Grangier},\ and\ \citenamefont
  {Browaeys}}]{Wilk2010}%
  \BibitemOpen
  \bibfield  {author} {\bibinfo {author} {\bibfnamefont {T.}~\bibnamefont
  {Wilk}}, \bibinfo {author} {\bibfnamefont {A.}~\bibnamefont {Ga\"etan}},
  \bibinfo {author} {\bibfnamefont {C.}~\bibnamefont {Evellin}}, \bibinfo
  {author} {\bibfnamefont {J.}~\bibnamefont {Wolters}}, \bibinfo {author}
  {\bibfnamefont {Y.}~\bibnamefont {Miroshnychenko}}, \bibinfo {author}
  {\bibfnamefont {P.}~\bibnamefont {Grangier}}, \ and\ \bibinfo {author}
  {\bibfnamefont {A.}~\bibnamefont {Browaeys}},\ }\bibfield  {title} {\enquote
  {\bibinfo {title} {{E}ntanglement of {T}wo {I}ndividual {N}eutral {A}toms
  {U}sing {R}ydberg {B}lockade},}\ }\href {\doibase 10.1103/PhysRevLett.104.010502} {\bibfield  {journal} {\bibinfo  {journal}
  {Phys. Rev. Lett.}\ }\textbf {\bibinfo {volume} {104}},\ \bibinfo {pages}
  {010502} (\bibinfo {year} {2010})}\BibitemShut {NoStop}%
\bibitem [{\citenamefont {Isenhower}\ \emph {et~al.}(2010)\citenamefont
  {Isenhower}, \citenamefont {Urban}, \citenamefont {Zhang}, \citenamefont
  {Gill}, \citenamefont {Henage}, \citenamefont {Johnson}, \citenamefont
  {Walker},\ and\ \citenamefont {Saffman}}]{Isenhower2010}%
  \BibitemOpen
  \bibfield  {author} {\bibinfo {author} {\bibfnamefont {L.}~\bibnamefont
  {Isenhower}}, \bibinfo {author} {\bibfnamefont {E.}~\bibnamefont {Urban}},
  \bibinfo {author} {\bibfnamefont {X.~L.}\ \bibnamefont {Zhang}}, \bibinfo
  {author} {\bibfnamefont {A.~T.}\ \bibnamefont {Gill}}, \bibinfo {author}
  {\bibfnamefont {T.}~\bibnamefont {Henage}}, \bibinfo {author} {\bibfnamefont
  {T.~A.}\ \bibnamefont {Johnson}}, \bibinfo {author} {\bibfnamefont {T.~G.}\
  \bibnamefont {Walker}}, \ and\ \bibinfo {author} {\bibfnamefont
  {M.}~\bibnamefont {Saffman}},\ }\bibfield  {title} {\enquote {\bibinfo
  {title} {{D}emonstration of a {N}eutral {A}tom {C}ontrolled-{N}{O}{T}
  {Q}uantum {G}ate},}\ }\href {\doibase 10.1103/PhysRevLett.104.010503}
  {\bibfield  {journal} {\bibinfo  {journal} {Phys. Rev. Lett.}\ }\textbf
  {\bibinfo {volume} {104}},\ \bibinfo {pages} {010503} (\bibinfo {year}
  {2010})}\BibitemShut {NoStop}%
\bibitem [{\citenamefont {Comparat}\ and\ \citenamefont
  {Pillet}(2010)}]{Comparat2010}%
  \BibitemOpen
  \bibfield  {author} {\bibinfo {author} {\bibfnamefont {D.}~\bibnamefont
  {Comparat}}\ and\ \bibinfo {author} {\bibfnamefont {P.}~\bibnamefont
  {Pillet}},\ }\bibfield  {title} {\enquote {\bibinfo {title} {{D}ipole
  blockade in a cold {R}ydberg atomic sample [{I}nvited]},}\ }\href {\doibase 10.1364/JOSAB.27.00A208} {\bibfield  {journal} {\bibinfo  {journal} {J. Opt.
  Soc. Am. B}\ }\textbf {\bibinfo {volume} {27}},\ \bibinfo {pages} {A208}
  (\bibinfo {year} {2010})}\BibitemShut {NoStop}%
\bibitem [{\citenamefont {Barredo}\ \emph {et~al.}(2014)\citenamefont
  {Barredo}, \citenamefont {Ravets}, \citenamefont {Labuhn}, \citenamefont
  {B\'eguin}, \citenamefont {Vernier}, \citenamefont {Nogrette}, \citenamefont
  {Lahaye},\ and\ \citenamefont {Browaeys}}]{Barredo2014}%
  \BibitemOpen
  \bibfield  {author} {\bibinfo {author} {\bibfnamefont {D.}~\bibnamefont
  {Barredo}}, \bibinfo {author} {\bibfnamefont {S.}~\bibnamefont {Ravets}},
  \bibinfo {author} {\bibfnamefont {H.}~\bibnamefont {Labuhn}}, \bibinfo
  {author} {\bibfnamefont {L.}~\bibnamefont {B\'eguin}}, \bibinfo {author}
  {\bibfnamefont {A.}~\bibnamefont {Vernier}}, \bibinfo {author} {\bibfnamefont
  {F.}~\bibnamefont {Nogrette}}, \bibinfo {author} {\bibfnamefont
  {T.}~\bibnamefont {Lahaye}}, \ and\ \bibinfo {author} {\bibfnamefont
  {A.}~\bibnamefont {Browaeys}},\ }\bibfield  {title} {\enquote {\bibinfo
  {title} {{D}emonstration of a {S}trong {R}ydberg {B}lockade in {T}hree-{A}tom
  {S}ystems with {A}nisotropic {I}nteractions},}\ }\href {\doibase 10.1103/PhysRevLett.112.183002} {\bibfield  {journal} {\bibinfo  {journal}
  {Phys. Rev. Lett.}\ }\textbf {\bibinfo {volume} {112}},\ \bibinfo {pages}
  {183002} (\bibinfo {year} {2014})}\BibitemShut {NoStop}%
\bibitem [{\citenamefont {Hankin}\ \emph {et~al.}(2014)\citenamefont {Hankin},
  \citenamefont {Jau}, \citenamefont {Parazzoli}, \citenamefont {Chou},
  \citenamefont {Armstrong}, \citenamefont {Landahl},\ and\ \citenamefont
  {Biedermann}}]{Hankin2014}%
  \BibitemOpen
  \bibfield  {author} {\bibinfo {author} {\bibfnamefont {A.~M.}\ \bibnamefont
  {Hankin}}, \bibinfo {author} {\bibfnamefont {Y.-Y.}\ \bibnamefont {Jau}},
  \bibinfo {author} {\bibfnamefont {L.~P.}\ \bibnamefont {Parazzoli}}, \bibinfo
  {author} {\bibfnamefont {C.~W.}\ \bibnamefont {Chou}}, \bibinfo {author}
  {\bibfnamefont {D.~J.}\ \bibnamefont {Armstrong}}, \bibinfo {author}
  {\bibfnamefont {A.~J.}\ \bibnamefont {Landahl}}, \ and\ \bibinfo {author}
  {\bibfnamefont {G.~W.}\ \bibnamefont {Biedermann}},\ }\bibfield  {title}
  {\enquote {\bibinfo {title} {{T}wo-atom {R}ydberg blockade using direct 6$s$
  to $np$ excitation},}\ }\href {\doibase 10.1103/PhysRevA.89.033416}
  {\bibfield  {journal} {\bibinfo  {journal} {Phys. Rev. A}\ }\textbf {\bibinfo
  {volume} {89}},\ \bibinfo {pages} {033416} (\bibinfo {year}
  {2014})}\BibitemShut {NoStop}%
\bibitem [{\citenamefont {Jau}\ \emph {et~al.}(2015)\citenamefont {Jau},
  \citenamefont {Hankin}, \citenamefont {Keating}, \citenamefont {Deutsch},\
  and\ \citenamefont {Biedermann}}]{Jau2015}%
  \BibitemOpen
  \bibfield  {author} {\bibinfo {author} {\bibfnamefont {Y.-Y.}\ \bibnamefont
  {Jau}}, \bibinfo {author} {\bibfnamefont {A.~M.}\ \bibnamefont {Hankin}},
  \bibinfo {author} {\bibfnamefont {T.}~\bibnamefont {Keating}}, \bibinfo
  {author} {\bibfnamefont {I.~H.}\ \bibnamefont {Deutsch}}, \ and\ \bibinfo
  {author} {\bibfnamefont {G.~W.}\ \bibnamefont {Biedermann}},\ }\bibfield
  {title} {\enquote {\bibinfo {title} {Entangling atomic spins with a strong
  rydberg-dressed interaction},}\ }\href {http://arxiv.org/abs/1501.03862}
  {\bibfield  {journal} {\bibinfo  {journal} {arXiv:1501.03862}\ } (\bibinfo
  {year} {2015})}\BibitemShut {NoStop}%
\bibitem [{\citenamefont {Anderson}\ \emph {et~al.}(1998)\citenamefont
  {Anderson}, \citenamefont {Veale},\ and\ \citenamefont
  {Gallagher}}]{Anderson1998}%
  \BibitemOpen
  \bibfield  {author} {\bibinfo {author} {\bibfnamefont {W.~R.}\ \bibnamefont
  {Anderson}}, \bibinfo {author} {\bibfnamefont {J.~R.}\ \bibnamefont {Veale}},
  \ and\ \bibinfo {author} {\bibfnamefont {T.~F.}\ \bibnamefont {Gallagher}},\
  }\bibfield  {title} {\enquote {\bibinfo {title} {{R}esonant {D}ipole-{D}ipole
  {E}nergy {T}ransfer in a {N}early {F}rozen {R}ydberg {G}as},}\ }\href {\doibase 10.1103/PhysRevLett.80.249} {\bibfield  {journal} {\bibinfo
  {journal} {Phys. Rev. Lett.}\ }\textbf {\bibinfo {volume} {80}},\ \bibinfo
  {pages} {249} (\bibinfo {year} {1998})}\BibitemShut {NoStop}%
\bibitem [{\citenamefont {Mourachko}\ \emph {et~al.}(1998)\citenamefont
  {Mourachko}, \citenamefont {Comparat}, \citenamefont {de~Tomasi},
  \citenamefont {Fioretti}, \citenamefont {Nosbaum}, \citenamefont {Akulin},\
  and\ \citenamefont {Pillet}}]{Mourachko1998}%
  \BibitemOpen
  \bibfield  {author} {\bibinfo {author} {\bibfnamefont {I.}~\bibnamefont
  {Mourachko}}, \bibinfo {author} {\bibfnamefont {D.}~\bibnamefont {Comparat}},
  \bibinfo {author} {\bibfnamefont {F.}~\bibnamefont {de~Tomasi}}, \bibinfo
  {author} {\bibfnamefont {A.}~\bibnamefont {Fioretti}}, \bibinfo {author}
  {\bibfnamefont {P.}~\bibnamefont {Nosbaum}}, \bibinfo {author} {\bibfnamefont
  {V.~M.}\ \bibnamefont {Akulin}}, \ and\ \bibinfo {author} {\bibfnamefont
  {P.}~\bibnamefont {Pillet}},\ }\bibfield  {title} {\enquote {\bibinfo {title}
  {{M}any-{B}ody {E}ffects in a {F}rozen {R}ydberg {G}as},}\ }\href {\doibase 10.1103/PhysRevLett.80.253} {\bibfield  {journal} {\bibinfo  {journal} {Phys.
  Rev. Lett.}\ }\textbf {\bibinfo {volume} {80}},\ \bibinfo {pages} {253}
  (\bibinfo {year} {1998})}\BibitemShut {NoStop}%
\bibitem [{\citenamefont {G\"{u}nter}\ \emph {et~al.}(2013)\citenamefont
  {G\"{u}nter}, \citenamefont {Schempp}, \citenamefont
  {{Robert-de-Saint-Vincent}}, \citenamefont {Gavryusev}, \citenamefont
  {Helmrich}, \citenamefont {Hofmann}, \citenamefont {Whitlock},\ and\
  \citenamefont {Weidem\"{u}ller}}]{Gunter2013}%
  \BibitemOpen
  \bibfield  {author} {\bibinfo {author} {\bibfnamefont {G.}~\bibnamefont
  {G\"{u}nter}}, \bibinfo {author} {\bibfnamefont {H.}~\bibnamefont {Schempp}},
  \bibinfo {author} {\bibfnamefont {M.}~\bibnamefont
  {{Robert-de-Saint-Vincent}}}, \bibinfo {author} {\bibfnamefont
  {V.}~\bibnamefont {Gavryusev}}, \bibinfo {author} {\bibfnamefont
  {S.}~\bibnamefont {Helmrich}}, \bibinfo {author} {\bibfnamefont {C.~S.}\
  \bibnamefont {Hofmann}}, \bibinfo {author} {\bibfnamefont {S.}~\bibnamefont
  {Whitlock}}, \ and\ \bibinfo {author} {\bibfnamefont {M.}~\bibnamefont
  {Weidem\"{u}ller}},\ }\bibfield  {title} {\enquote {\bibinfo {title}
  {{O}bserving the {D}ynamics of {D}ipole-{M}ediated {E}nergy {T}ransport by
  {I}nteraction-{E}nhanced {I}maging},}\ }\href {\doibase 10.1126/science.1244843} {\bibfield  {journal} {\bibinfo  {journal}
  {Science}\ }\textbf {\bibinfo {volume} {342}},\ \bibinfo {pages} {954}
  (\bibinfo {year} {2013})}\BibitemShut {NoStop}%
\bibitem [{\citenamefont {Barredo}\ \emph {et~al.}(2015)\citenamefont
  {Barredo}, \citenamefont {Labuhn}, \citenamefont {Ravets}, \citenamefont
  {Lahaye}, \citenamefont {Browaeys},\ and\ \citenamefont
  {Adams}}]{Barredo2014b}%
  \BibitemOpen
  \bibfield  {author} {\bibinfo {author} {\bibfnamefont {D.}~\bibnamefont
  {Barredo}}, \bibinfo {author} {\bibfnamefont {H.}~\bibnamefont {Labuhn}},
  \bibinfo {author} {\bibfnamefont {S.}~\bibnamefont {Ravets}}, \bibinfo
  {author} {\bibfnamefont {T.}~\bibnamefont {Lahaye}}, \bibinfo {author}
  {\bibfnamefont {A.}~\bibnamefont {Browaeys}}, \ and\ \bibinfo {author}
  {\bibfnamefont {C.~S.}\ \bibnamefont {Adams}},\ }\bibfield  {title} {\enquote
  {\bibinfo {title} {Coherent excitation transfer in a spin chain of three
  rydberg atoms},}\ }\href {\doibase 10.1103/PhysRevLett.114.113002} {\bibfield
   {journal} {\bibinfo  {journal} {Phys. Rev. Lett.}\ }\textbf {\bibinfo
  {volume} {114}},\ \bibinfo {pages} {113002} (\bibinfo {year}
  {2015})}\BibitemShut {NoStop}%
\bibitem [{\citenamefont {Reinhard}\ \emph {et~al.}(2007)\citenamefont
  {Reinhard}, \citenamefont {Liebisch}, \citenamefont {Knuffman},\ and\
  \citenamefont {Raithel}}]{Reinhard2007}%
  \BibitemOpen
  \bibfield  {author} {\bibinfo {author} {\bibfnamefont {A.}~\bibnamefont
  {Reinhard}}, \bibinfo {author} {\bibfnamefont {T.~C.}\ \bibnamefont
  {Liebisch}}, \bibinfo {author} {\bibfnamefont {B.}~\bibnamefont {Knuffman}},
  \ and\ \bibinfo {author} {\bibfnamefont {G.}~\bibnamefont {Raithel}},\
  }\bibfield  {title} {\enquote {\bibinfo {title} {{L}evel shifts of rubidium
  {R}ydberg states due to binary interactions},}\ }\href {\doibase 10.1103/PhysRevA.75.032712} {\bibfield  {journal} {\bibinfo  {journal} {Phys.
  Rev. A}\ }\textbf {\bibinfo {volume} {75}},\ \bibinfo {pages} {032712}
  (\bibinfo {year} {2007})}\BibitemShut {NoStop}%
\bibitem [{\citenamefont {Walker}\ and\ \citenamefont
  {Saffman}(2008)}]{Walker2008}%
  \BibitemOpen
  \bibfield  {author} {\bibinfo {author} {\bibfnamefont {T.~G.}\ \bibnamefont
  {Walker}}\ and\ \bibinfo {author} {\bibfnamefont {M.}~\bibnamefont
  {Saffman}},\ }\bibfield  {title} {\enquote {\bibinfo {title} {{C}onsequences
  of {Z}eeman degeneracy for the van der {W}aals blockade between {R}ydberg
  atoms},}\ }\href {\doibase 10.1103/PhysRevA.77.032723} {\bibfield  {journal}
  {\bibinfo  {journal} {Phys. Rev. A}\ }\textbf {\bibinfo {volume} {77}},\
  \bibinfo {pages} {032723} (\bibinfo {year} {2008})}\BibitemShut {NoStop}%
\bibitem [{\citenamefont {Marcassa}\ and\ \citenamefont
  {Shaffer}(2014)}]{Marcassa2014}%
  \BibitemOpen
  \bibfield  {author} {\bibinfo {author} {\bibfnamefont {L.~G.}\ \bibnamefont
  {Marcassa}}\ and\ \bibinfo {author} {\bibfnamefont {J.~P.}\ \bibnamefont
  {Shaffer}},\ }\bibfield  {title} {\enquote {\bibinfo {title} {Chapter two -
  interactions in ultracold rydberg gases},}\ }in\ \href {\doibase 10.1016/B978-0-12-800129-5.00002-X} {\emph {\bibinfo {booktitle} {Advances In
  Atomic, Molecular, and Optical Physics}}},\ Vol.~\bibinfo {volume} {63},\
  \bibinfo {editor} {edited by\ \bibinfo {editor} {\bibfnamefont
  {E.}~\bibnamefont {Arimondo}}, \bibinfo {editor} {\bibfnamefont {P.~R.}\
  \bibnamefont {Berman}}, \ and\ \bibinfo {editor} {\bibfnamefont {C.~C.}\
  \bibnamefont {Lin}}}\ (\bibinfo  {publisher} {Academic Press},\ \bibinfo
  {year} {2014})\ pp.\ \bibinfo {pages} {47 -- 133}\BibitemShut {NoStop}%
\bibitem [{\citenamefont {Vermersch}\ \emph {et~al.}(2015)\citenamefont
  {Vermersch}, \citenamefont {Glaetzle},\ and\ \citenamefont
  {Zoller}}]{Vermersch2015}%
  \BibitemOpen
  \bibfield  {author} {\bibinfo {author} {\bibfnamefont {B.}~\bibnamefont
  {Vermersch}}, \bibinfo {author} {\bibfnamefont {A.~W.}\ \bibnamefont
  {Glaetzle}}, \ and\ \bibinfo {author} {\bibfnamefont {P.}~\bibnamefont
  {Zoller}},\ }\bibfield  {title} {\enquote {\bibinfo {title} {{M}agic
  distances in the blockade mechanism of {R}ydberg $p$ and $d$ states},}\
  }\href {\doibase 10.1103/PhysRevA.91.023411} {\bibfield  {journal} {\bibinfo
  {journal} {Phys. Rev. A}\ }\textbf {\bibinfo {volume} {91}},\ \bibinfo
  {pages} {023411} (\bibinfo {year} {2015})}\BibitemShut {NoStop}%
\bibitem [{\citenamefont {Slichter}(1990)}]{Slichter1990}%
  \BibitemOpen
  \bibfield  {author} {\bibinfo {author} {\bibfnamefont {C.}~\bibnamefont
  {Slichter}},\ }\href {\doibase 10.1007/978-3-662-09441-9} {\emph {\bibinfo
  {title} {{P}rinciples of {M}agnetic {R}esonance}}},\ Springer Series in
  Solid-State Sciences\ (\bibinfo  {publisher} {Springer Berlin Heidelberg},\
  \bibinfo {year} {1990})\BibitemShut {NoStop}%
\bibitem [{\citenamefont {Ni}\ \emph {et~al.}(2009)\citenamefont {Ni},
  \citenamefont {Ospelkaus}, \citenamefont {Nesbitt}, \citenamefont {Ye},\ and\
  \citenamefont {Jin}}]{Ni2009}%
  \BibitemOpen
  \bibfield  {author} {\bibinfo {author} {\bibfnamefont {K.-K.}\ \bibnamefont
  {Ni}}, \bibinfo {author} {\bibfnamefont {S.}~\bibnamefont {Ospelkaus}},
  \bibinfo {author} {\bibfnamefont {D.~J.}\ \bibnamefont {Nesbitt}}, \bibinfo
  {author} {\bibfnamefont {J.}~\bibnamefont {Ye}}, \ and\ \bibinfo {author}
  {\bibfnamefont {D.~S.}\ \bibnamefont {Jin}},\ }\bibfield  {title} {\enquote
  {\bibinfo {title} {{A} dipolar gas of ultracold molecules},}\ }\href {\doibase 10.1039/B911779B} {\bibfield  {journal} {\bibinfo  {journal} {Phys.
  Chem. Chem. Phys.}\ }\textbf {\bibinfo {volume} {11}},\ \bibinfo {pages}
  {9626} (\bibinfo {year} {2009})}\BibitemShut {NoStop}%
\bibitem [{\citenamefont {Hazzard}\ \emph {et~al.}(2014)\citenamefont
  {Hazzard}, \citenamefont {Gadway}, \citenamefont {Foss-Feig}, \citenamefont
  {Yan}, \citenamefont {Moses}, \citenamefont {Covey}, \citenamefont {Yao},
  \citenamefont {Lukin}, \citenamefont {Ye}, \citenamefont {Jin},\ and\
  \citenamefont {Rey}}]{Hazzard2014}%
  \BibitemOpen
  \bibfield  {author} {\bibinfo {author} {\bibfnamefont {K.~R.~A.}\
  \bibnamefont {Hazzard}}, \bibinfo {author} {\bibfnamefont {B.}~\bibnamefont
  {Gadway}}, \bibinfo {author} {\bibfnamefont {M.}~\bibnamefont {Foss-Feig}},
  \bibinfo {author} {\bibfnamefont {B.}~\bibnamefont {Yan}}, \bibinfo {author}
  {\bibfnamefont {S.~A.}\ \bibnamefont {Moses}}, \bibinfo {author}
  {\bibfnamefont {J.~P.}\ \bibnamefont {Covey}}, \bibinfo {author}
  {\bibfnamefont {N.~Y.}\ \bibnamefont {Yao}}, \bibinfo {author} {\bibfnamefont
  {M.~D.}\ \bibnamefont {Lukin}}, \bibinfo {author} {\bibfnamefont
  {J.}~\bibnamefont {Ye}}, \bibinfo {author} {\bibfnamefont {D.~S.}\
  \bibnamefont {Jin}}, \ and\ \bibinfo {author} {\bibfnamefont {A.~M.}\
  \bibnamefont {Rey}},\ }\bibfield  {title} {\enquote {\bibinfo {title}
  {{M}any-{B}ody {D}ynamics of {D}ipolar {M}olecules in an {O}ptical
  {L}attice},}\ }\href {\doibase 10.1103/PhysRevLett.113.195302} {\bibfield
  {journal} {\bibinfo  {journal} {Phys. Rev. Lett.}\ }\textbf {\bibinfo
  {volume} {113}},\ \bibinfo {pages} {195302} (\bibinfo {year}
  {2014})}\BibitemShut {NoStop}%
\bibitem [{\citenamefont {Lahaye}\ \emph {et~al.}(2009)\citenamefont {Lahaye},
  \citenamefont {Menotti}, \citenamefont {Santos}, \citenamefont {Lewenstein},\
  and\ \citenamefont {Pfau}}]{Lahaye2009}%
  \BibitemOpen
  \bibfield  {author} {\bibinfo {author} {\bibfnamefont {T.}~\bibnamefont
  {Lahaye}}, \bibinfo {author} {\bibfnamefont {C.}~\bibnamefont {Menotti}},
  \bibinfo {author} {\bibfnamefont {L.}~\bibnamefont {Santos}}, \bibinfo
  {author} {\bibfnamefont {M.}~\bibnamefont {Lewenstein}}, \ and\ \bibinfo
  {author} {\bibfnamefont {T.}~\bibnamefont {Pfau}},\ }\bibfield  {title}
  {\enquote {\bibinfo {title} {{T}he physics of dipolar bosonic quantum
  gases},}\ }\href {http://stacks.iop.org/0034-4885/72/i=12/a=126401}
  {\bibfield  {journal} {\bibinfo  {journal} {Rep. Prog. Phys.}\ }\textbf
  {\bibinfo {volume} {72}},\ \bibinfo {pages} {126401} (\bibinfo {year}
  {2009})}\BibitemShut {NoStop}%
\bibitem [{\citenamefont {Carroll}\ \emph {et~al.}(2004)\citenamefont
  {Carroll}, \citenamefont {Claringbould}, \citenamefont {Goodsell},
  \citenamefont {Lim},\ and\ \citenamefont {Noel}}]{Carroll2004}%
  \BibitemOpen
  \bibfield  {author} {\bibinfo {author} {\bibfnamefont {T.~J.}\ \bibnamefont
  {Carroll}}, \bibinfo {author} {\bibfnamefont {K.}~\bibnamefont
  {Claringbould}}, \bibinfo {author} {\bibfnamefont {A.}~\bibnamefont
  {Goodsell}}, \bibinfo {author} {\bibfnamefont {M.~J.}\ \bibnamefont {Lim}}, \
  and\ \bibinfo {author} {\bibfnamefont {M.~W.}\ \bibnamefont {Noel}},\
  }\bibfield  {title} {\enquote {\bibinfo {title} {{A}ngular {D}ependence of
  the {D}ipole-{D}ipole {I}nteraction in a {N}early {O}ne-{D}imensional
  {S}ample of {R}ydberg {A}toms},}\ }\href {\doibase 10.1103/PhysRevLett.93.153001} {\bibfield  {journal} {\bibinfo  {journal}
  {Phys. Rev. Lett.}\ }\textbf {\bibinfo {volume} {93}},\ \bibinfo {pages}
  {153001} (\bibinfo {year} {2004})}\BibitemShut {NoStop}%
\bibitem [{\citenamefont {Cabral}\ \emph {et~al.}(2011)\citenamefont {Cabral},
  \citenamefont {Kondo}, \citenamefont {Gon{\c c}alves}, \citenamefont
  {Nascimento}, \citenamefont {Marcassa}, \citenamefont {Booth}, \citenamefont
  {Tallant}, \citenamefont {Schwettmann}, \citenamefont {Overstreet},
  \citenamefont {Sedlacek},\ and\ \citenamefont {Shaffer}}]{Cabral2011}%
  \BibitemOpen
  \bibfield  {author} {\bibinfo {author} {\bibfnamefont {J.~S.}\ \bibnamefont
  {Cabral}}, \bibinfo {author} {\bibfnamefont {J.~M.}\ \bibnamefont {Kondo}},
  \bibinfo {author} {\bibfnamefont {L.~F.}\ \bibnamefont {Gon{\c c}alves}},
  \bibinfo {author} {\bibfnamefont {V.~A.}\ \bibnamefont {Nascimento}},
  \bibinfo {author} {\bibfnamefont {L.~G.}\ \bibnamefont {Marcassa}}, \bibinfo
  {author} {\bibfnamefont {D.}~\bibnamefont {Booth}}, \bibinfo {author}
  {\bibfnamefont {J.}~\bibnamefont {Tallant}}, \bibinfo {author} {\bibfnamefont
  {A.}~\bibnamefont {Schwettmann}}, \bibinfo {author} {\bibfnamefont {K.~R.}\
  \bibnamefont {Overstreet}}, \bibinfo {author} {\bibfnamefont
  {J.}~\bibnamefont {Sedlacek}}, \ and\ \bibinfo {author} {\bibfnamefont
  {J.~P.}\ \bibnamefont {Shaffer}},\ }\bibfield  {title} {\enquote {\bibinfo
  {title} {{E}ffects of electric fields on ultracold {R}ydberg atom
  interactions},}\ }\href {http://stacks.iop.org/0953-4075/44/i=18/a=184007}
  {\bibfield  {journal} {\bibinfo  {journal} {J. Phys. B}\ }\textbf {\bibinfo
  {volume} {44}},\ \bibinfo {pages} {184007} (\bibinfo {year}
  {2011})}\BibitemShut {NoStop}%
\bibitem [{\citenamefont {Ravets}\ \emph {et~al.}(2014)\citenamefont {Ravets},
  \citenamefont {Labuhn}, \citenamefont {Barredo}, \citenamefont {Beguin},
  \citenamefont {Lahaye},\ and\ \citenamefont {Browaeys}}]{Ravets2014}%
  \BibitemOpen
  \bibfield  {author} {\bibinfo {author} {\bibfnamefont {S.}~\bibnamefont
  {Ravets}}, \bibinfo {author} {\bibfnamefont {H.}~\bibnamefont {Labuhn}},
  \bibinfo {author} {\bibfnamefont {D.}~\bibnamefont {Barredo}}, \bibinfo
  {author} {\bibfnamefont {L.}~\bibnamefont {Beguin}}, \bibinfo {author}
  {\bibfnamefont {T.}~\bibnamefont {Lahaye}}, \ and\ \bibinfo {author}
  {\bibfnamefont {A.}~\bibnamefont {Browaeys}},\ }\bibfield  {title} {\enquote
  {\bibinfo {title} {{C}oherent dipole-dipole coupling between two single
  {R}ydberg atoms at an electrically-tuned {F}{\"o}rster resonance},}\ }\href {http://dx.doi.org/10.1038/nphys3119} {\bibfield  {journal} {\bibinfo
  {journal} {Nat. Phys.}\ }\textbf {\bibinfo {volume} {10}},\ \bibinfo {pages}
  {914} (\bibinfo {year} {2014})}\BibitemShut {NoStop}%
\bibitem [{\citenamefont {Walker}\ and\ \citenamefont
  {Saffman}(2005)}]{Walker2005}%
  \BibitemOpen
  \bibfield  {author} {\bibinfo {author} {\bibfnamefont {T.~G.}\ \bibnamefont
  {Walker}}\ and\ \bibinfo {author} {\bibfnamefont {M.}~\bibnamefont
  {Saffman}},\ }\bibfield  {title} {\enquote {\bibinfo {title} {{Z}eros of
  {R}ydberg--{R}ydberg {F}\"oster interactions},}\ }\href {http://stacks.iop.org/0953-4075/38/i=2/a=022} {\bibfield  {journal}
  {\bibinfo  {journal} {J. Phys. B: At. Mol. Opt. Phys.}\ }\textbf {\bibinfo
  {volume} {38}},\ \bibinfo {pages} {S309} (\bibinfo {year}
  {2005})}\BibitemShut {NoStop}%
\bibitem [{\citenamefont {Anderson}\ \emph {et~al.}(2002)\citenamefont
  {Anderson}, \citenamefont {Robinson}, \citenamefont {Martin},\ and\
  \citenamefont {Gallagher}}]{Anderson2002}%
  \BibitemOpen
  \bibfield  {author} {\bibinfo {author} {\bibfnamefont {W.~R.}\ \bibnamefont
  {Anderson}}, \bibinfo {author} {\bibfnamefont {M.~P.}\ \bibnamefont
  {Robinson}}, \bibinfo {author} {\bibfnamefont {J.~D.~D.}\ \bibnamefont
  {Martin}}, \ and\ \bibinfo {author} {\bibfnamefont {T.~F.}\ \bibnamefont
  {Gallagher}},\ }\bibfield  {title} {\enquote {\bibinfo {title} {{D}ephasing
  of resonant energy transfer in a cold {R}ydberg gas},}\ }\href {\doibase 10.1103/PhysRevA.65.063404} {\bibfield  {journal} {\bibinfo  {journal} {Phys.
  Rev. A}\ }\textbf {\bibinfo {volume} {65}},\ \bibinfo {pages} {063404}
  (\bibinfo {year} {2002})}\BibitemShut {NoStop}%
\bibitem [{\citenamefont {Mudrich}\ \emph {et~al.}(2005)\citenamefont
  {Mudrich}, \citenamefont {Zahzam}, \citenamefont {Vogt}, \citenamefont
  {Comparat},\ and\ \citenamefont {Pillet}}]{Mudrich2005}%
  \BibitemOpen
  \bibfield  {author} {\bibinfo {author} {\bibfnamefont {M.}~\bibnamefont
  {Mudrich}}, \bibinfo {author} {\bibfnamefont {N.}~\bibnamefont {Zahzam}},
  \bibinfo {author} {\bibfnamefont {T.}~\bibnamefont {Vogt}}, \bibinfo {author}
  {\bibfnamefont {D.}~\bibnamefont {Comparat}}, \ and\ \bibinfo {author}
  {\bibfnamefont {P.}~\bibnamefont {Pillet}},\ }\bibfield  {title} {\enquote
  {\bibinfo {title} {{B}ack and {F}orth {T}ransfer and {C}oherent {C}oupling in
  a {C}old {R}ydberg {D}ipole {G}as},}\ }\href {\doibase 10.1103/PhysRevLett.95.233002} {\bibfield  {journal} {\bibinfo  {journal}
  {Phys. Rev. Lett.}\ }\textbf {\bibinfo {volume} {95}},\ \bibinfo {pages}
  {233002} (\bibinfo {year} {2005})}\BibitemShut {NoStop}%
\bibitem [{\citenamefont {Vogt}\ \emph {et~al.}(2007)\citenamefont {Vogt},
  \citenamefont {Viteau}, \citenamefont {Chotia}, \citenamefont {Zhao},
  \citenamefont {Comparat},\ and\ \citenamefont {Pillet}}]{Vogt2007}%
  \BibitemOpen
  \bibfield  {author} {\bibinfo {author} {\bibfnamefont {T.}~\bibnamefont
  {Vogt}}, \bibinfo {author} {\bibfnamefont {M.}~\bibnamefont {Viteau}},
  \bibinfo {author} {\bibfnamefont {A.}~\bibnamefont {Chotia}}, \bibinfo
  {author} {\bibfnamefont {J.}~\bibnamefont {Zhao}}, \bibinfo {author}
  {\bibfnamefont {D.}~\bibnamefont {Comparat}}, \ and\ \bibinfo {author}
  {\bibfnamefont {P.}~\bibnamefont {Pillet}},\ }\bibfield  {title} {\enquote
  {\bibinfo {title} {{E}lectric-{F}ield {I}nduced {D}ipole {B}lockade with
  {R}ydberg {A}toms},}\ }\href {\doibase 10.1103/PhysRevLett.99.073002}
  {\bibfield  {journal} {\bibinfo  {journal} {Phys. Rev. Lett.}\ }\textbf
  {\bibinfo {volume} {99}},\ \bibinfo {pages} {073002} (\bibinfo {year}
  {2007})}\BibitemShut {NoStop}%
\bibitem [{\citenamefont {{van Ditzhuijzen}}\ \emph {et~al.}(2008)\citenamefont
  {{van Ditzhuijzen}}, \citenamefont {Koenderink}, \citenamefont {Hern\'andez},
  \citenamefont {Robicheaux}, \citenamefont {Noordam},\ and\ \citenamefont
  {{van Linden van den Heuvell}}}]{Ditzhuijzen2008}%
  \BibitemOpen
  \bibfield  {author} {\bibinfo {author} {\bibfnamefont {C.~S.~E.}\
  \bibnamefont {{van Ditzhuijzen}}}, \bibinfo {author} {\bibfnamefont {A.~F.}\
  \bibnamefont {Koenderink}}, \bibinfo {author} {\bibfnamefont {J.~V.}\
  \bibnamefont {Hern\'andez}}, \bibinfo {author} {\bibfnamefont
  {F.}~\bibnamefont {Robicheaux}}, \bibinfo {author} {\bibfnamefont {L.~D.}\
  \bibnamefont {Noordam}}, \ and\ \bibinfo {author} {\bibfnamefont {H.~B.}\
  \bibnamefont {{van Linden van den Heuvell}}},\ }\bibfield  {title} {\enquote
  {\bibinfo {title} {{S}patially {R}esolved {O}bservation of {D}ipole-{D}ipole
  {I}nteraction between {R}ydberg {A}toms},}\ }\href {\doibase 10.1103/PhysRevLett.100.243201} {\bibfield  {journal} {\bibinfo  {journal}
  {Phys. Rev. Lett.}\ }\textbf {\bibinfo {volume} {100}},\ \bibinfo {pages}
  {243201} (\bibinfo {year} {2008})}\BibitemShut {NoStop}%
\bibitem [{\citenamefont {Ryabtsev}\ \emph {et~al.}(2010)\citenamefont
  {Ryabtsev}, \citenamefont {Tretyakov}, \citenamefont {Beterov},\ and\
  \citenamefont {Entin}}]{Ryabtsev2010}%
  \BibitemOpen
  \bibfield  {author} {\bibinfo {author} {\bibfnamefont {I.~I.}\ \bibnamefont
  {Ryabtsev}}, \bibinfo {author} {\bibfnamefont {D.~B.}\ \bibnamefont
  {Tretyakov}}, \bibinfo {author} {\bibfnamefont {I.~I.}\ \bibnamefont
  {Beterov}}, \ and\ \bibinfo {author} {\bibfnamefont {V.~M.}\ \bibnamefont
  {Entin}},\ }\bibfield  {title} {\enquote {\bibinfo {title} {{O}bservation of
  the {S}tark-{T}uned {F}\"orster {R}esonance between {T}wo {R}ydberg
  {A}toms},}\ }\href {\doibase 10.1103/PhysRevLett.104.073003} {\bibfield
  {journal} {\bibinfo  {journal} {Phys. Rev. Lett.}\ }\textbf {\bibinfo
  {volume} {104}},\ \bibinfo {pages} {073003} (\bibinfo {year}
  {2010})}\BibitemShut {NoStop}%
\bibitem [{\citenamefont {Nipper}\ \emph
  {et~al.}(2012{\natexlab{a}})\citenamefont {Nipper}, \citenamefont {Balewski},
  \citenamefont {Krupp}, \citenamefont {Butscher}, \citenamefont {L\"ow},\ and\
  \citenamefont {Pfau}}]{Nipper2012}%
  \BibitemOpen
  \bibfield  {author} {\bibinfo {author} {\bibfnamefont {J.}~\bibnamefont
  {Nipper}}, \bibinfo {author} {\bibfnamefont {J.~B.}\ \bibnamefont
  {Balewski}}, \bibinfo {author} {\bibfnamefont {A.~T.}\ \bibnamefont {Krupp}},
  \bibinfo {author} {\bibfnamefont {B.}~\bibnamefont {Butscher}}, \bibinfo
  {author} {\bibfnamefont {R.}~\bibnamefont {L\"ow}}, \ and\ \bibinfo {author}
  {\bibfnamefont {T.}~\bibnamefont {Pfau}},\ }\bibfield  {title} {\enquote
  {\bibinfo {title} {{H}ighly {R}esolved {M}easurements of {S}tark-{T}uned
  {F}\"orster {R}esonances between {R}ydberg {A}toms},}\ }\href {\doibase 10.1103/PhysRevLett.108.113001} {\bibfield  {journal} {\bibinfo  {journal}
  {Phys. Rev. Lett.}\ }\textbf {\bibinfo {volume} {108}},\ \bibinfo {pages}
  {113001} (\bibinfo {year} {2012}{\natexlab{a}})}\BibitemShut {NoStop}%
\bibitem [{\citenamefont {Nipper}\ \emph
  {et~al.}(2012{\natexlab{b}})\citenamefont {Nipper}, \citenamefont {Balewski},
  \citenamefont {Krupp}, \citenamefont {Hofferberth}, \citenamefont {L\"ow},\
  and\ \citenamefont {Pfau}}]{Nipper2012b}%
  \BibitemOpen
  \bibfield  {author} {\bibinfo {author} {\bibfnamefont {J.}~\bibnamefont
  {Nipper}}, \bibinfo {author} {\bibfnamefont {J.~B.}\ \bibnamefont
  {Balewski}}, \bibinfo {author} {\bibfnamefont {A.~T.}\ \bibnamefont {Krupp}},
  \bibinfo {author} {\bibfnamefont {S.}~\bibnamefont {Hofferberth}}, \bibinfo
  {author} {\bibfnamefont {R.}~\bibnamefont {L\"ow}}, \ and\ \bibinfo {author}
  {\bibfnamefont {T.}~\bibnamefont {Pfau}},\ }\bibfield  {title} {\enquote
  {\bibinfo {title} {{A}tomic {P}air-{S}tate {I}nterferometer: {C}ontrolling
  and {M}easuring an {I}nteraction-{I}nduced {P}hase {S}hift in
  {R}ydberg-{A}tom {P}airs},}\ }\href {\doibase 10.1103/PhysRevX.2.031011}
  {\bibfield  {journal} {\bibinfo  {journal} {Phys. Rev. X}\ }\textbf {\bibinfo
  {volume} {2}},\ \bibinfo {pages} {031011} (\bibinfo {year}
  {2012}{\natexlab{b}})}\BibitemShut {NoStop}%
\bibitem [{Note1()}]{Note1}%
  \BibitemOpen
  \bibinfo {note} {The electric field required to reach the resonance is weak
  enough to be in a regime of induced dipoles, by opposition to rigid
  dipoles.}\BibitemShut {Stop}%
\bibitem [{\citenamefont {B\'eguin}\ \emph {et~al.}(2013)\citenamefont
  {B\'eguin}, \citenamefont {Vernier}, \citenamefont {Chicireanu},
  \citenamefont {Lahaye},\ and\ \citenamefont {Browaeys}}]{Beguin2013}%
  \BibitemOpen
  \bibfield  {author} {\bibinfo {author} {\bibfnamefont {L.}~\bibnamefont
  {B\'eguin}}, \bibinfo {author} {\bibfnamefont {A.}~\bibnamefont {Vernier}},
  \bibinfo {author} {\bibfnamefont {R.}~\bibnamefont {Chicireanu}}, \bibinfo
  {author} {\bibfnamefont {T.}~\bibnamefont {Lahaye}}, \ and\ \bibinfo {author}
  {\bibfnamefont {A.}~\bibnamefont {Browaeys}},\ }\bibfield  {title} {\enquote
  {\bibinfo {title} {{D}irect {M}easurement of the van der {W}aals
  {I}nteraction between {T}wo {R}ydberg {A}toms},}\ }\href {\doibase 10.1103/PhysRevLett.110.263201} {\bibfield  {journal} {\bibinfo  {journal}
  {Phys. Rev. Lett.}\ }\textbf {\bibinfo {volume} {110}},\ \bibinfo {pages}
  {263201} (\bibinfo {year} {2013})}\BibitemShut {NoStop}%
\bibitem [{\citenamefont {Nogrette}\ \emph {et~al.}(2014)\citenamefont
  {Nogrette}, \citenamefont {Labuhn}, \citenamefont {Ravets}, \citenamefont
  {Barredo}, \citenamefont {B\'eguin}, \citenamefont {Vernier}, \citenamefont
  {Lahaye},\ and\ \citenamefont {Browaeys}}]{Nogrette2014}%
  \BibitemOpen
  \bibfield  {author} {\bibinfo {author} {\bibfnamefont {F.}~\bibnamefont
  {Nogrette}}, \bibinfo {author} {\bibfnamefont {H.}~\bibnamefont {Labuhn}},
  \bibinfo {author} {\bibfnamefont {S.}~\bibnamefont {Ravets}}, \bibinfo
  {author} {\bibfnamefont {D.}~\bibnamefont {Barredo}}, \bibinfo {author}
  {\bibfnamefont {L.}~\bibnamefont {B\'eguin}}, \bibinfo {author}
  {\bibfnamefont {A.}~\bibnamefont {Vernier}}, \bibinfo {author} {\bibfnamefont
  {T.}~\bibnamefont {Lahaye}}, \ and\ \bibinfo {author} {\bibfnamefont
  {A.}~\bibnamefont {Browaeys}},\ }\bibfield  {title} {\enquote {\bibinfo
  {title} {{S}ingle-{A}tom {T}rapping in {H}olographic 2{D} {A}rrays of
  {M}icrotraps with {A}rbitrary {G}eometries},}\ }\href {\doibase 10.1103/PhysRevX.4.021034} {\bibfield  {journal} {\bibinfo  {journal} {Phys.
  Rev. X}\ }\textbf {\bibinfo {volume} {4}},\ \bibinfo {pages} {021034}
  (\bibinfo {year} {2014})}\BibitemShut {NoStop}%
\bibitem [{Note2()}]{Note2}%
  \BibitemOpen
  \bibinfo {note} {For the Stark map calculation of the states $\ket {f_i}$, we
  found it necessary to include a large number of states in the basis to
  achieve convergence. Note that, due to the proximity of the hydrogenic
  manifold, the $B=3.3~{\protect \rm G}$ magnetic field cannot be taken into
  account perturbatively in the calculation. Here, we thus used a basis
  containing 756~states, with $3 \leq L \leq 56$, $-3 \leq m_L \leq 3$ and
  $-1/2 \leq m_S \leq 1/2$.}\BibitemShut {Stop}%
\bibitem [{\citenamefont {Li}\ \emph {et~al.}(2003)\citenamefont {Li},
  \citenamefont {Mourachko}, \citenamefont {Noel},\ and\ \citenamefont
  {Gallagher}}]{Li2003}%
  \BibitemOpen
  \bibfield  {author} {\bibinfo {author} {\bibfnamefont {W.}~\bibnamefont
  {Li}}, \bibinfo {author} {\bibfnamefont {I.}~\bibnamefont {Mourachko}},
  \bibinfo {author} {\bibfnamefont {M.~W.}\ \bibnamefont {Noel}}, \ and\
  \bibinfo {author} {\bibfnamefont {T.~F.}\ \bibnamefont {Gallagher}},\
  }\bibfield  {title} {\enquote {\bibinfo {title} {Millimeter-wave spectroscopy
  of cold rb rydberg atoms in a magneto-optical trap: Quantum defects of the
  \textit{ns} , \textit{np} , and \textit{nd} series},}\ }\href {\doibase 10.1103/PhysRevA.67.052502} {\bibfield  {journal} {\bibinfo  {journal} {Phys.
  Rev. A}\ }\textbf {\bibinfo {volume} {67}},\ \bibinfo {pages} {052502}
  (\bibinfo {year} {2003})}\BibitemShut {NoStop}%
\bibitem [{\citenamefont {Han}\ \emph {et~al.}(2006)\citenamefont {Han},
  \citenamefont {Jamil}, \citenamefont {Norum}, \citenamefont {Tanner},\ and\
  \citenamefont {Gallagher}}]{Han2006}%
  \BibitemOpen
  \bibfield  {author} {\bibinfo {author} {\bibfnamefont {J.}~\bibnamefont
  {Han}}, \bibinfo {author} {\bibfnamefont {Y.}~\bibnamefont {Jamil}}, \bibinfo
  {author} {\bibfnamefont {D.~V.~L.}\ \bibnamefont {Norum}}, \bibinfo {author}
  {\bibfnamefont {P.~J.}\ \bibnamefont {Tanner}}, \ and\ \bibinfo {author}
  {\bibfnamefont {T.~F.}\ \bibnamefont {Gallagher}},\ }\bibfield  {title}
  {\enquote {\bibinfo {title} {Rb $nf$ quantum defects from millimeter-wave
  spectroscopy of cold $^{85}\mathrm{Rb}$ rydberg atoms},}\ }\href {\doibase 10.1103/PhysRevA.74.054502} {\bibfield  {journal} {\bibinfo  {journal} {Phys.
  Rev. A}\ }\textbf {\bibinfo {volume} {74}},\ \bibinfo {pages} {054502}
  (\bibinfo {year} {2006})}\BibitemShut {NoStop}%
\bibitem [{\citenamefont {M\"uller}\ \emph {et~al.}(2009)\citenamefont
  {M\"uller}, \citenamefont {Lesanovsky}, \citenamefont {Weimer}, \citenamefont
  {B\"uchler},\ and\ \citenamefont {Zoller}}]{Muller2009}%
  \BibitemOpen
  \bibfield  {author} {\bibinfo {author} {\bibfnamefont {M.}~\bibnamefont
  {M\"uller}}, \bibinfo {author} {\bibfnamefont {I.}~\bibnamefont
  {Lesanovsky}}, \bibinfo {author} {\bibfnamefont {H.}~\bibnamefont {Weimer}},
  \bibinfo {author} {\bibfnamefont {H.~P.}\ \bibnamefont {B\"uchler}}, \ and\
  \bibinfo {author} {\bibfnamefont {P.}~\bibnamefont {Zoller}},\ }\bibfield
  {title} {\enquote {\bibinfo {title} {{M}esoscopic {R}ydberg {G}ate {B}ased on
  {E}lectromagnetically {I}nduced {T}ransparency},}\ }\href {\doibase 10.1103/PhysRevLett.102.170502} {\bibfield  {journal} {\bibinfo  {journal}
  {Phys. Rev. Lett.}\ }\textbf {\bibinfo {volume} {102}},\ \bibinfo {pages}
  {170502} (\bibinfo {year} {2009})}\BibitemShut {NoStop}%
\bibitem [{\citenamefont {Pohl}\ and\ \citenamefont {Berman}(2009)}]{Pohl2009}%
  \BibitemOpen
  \bibfield  {author} {\bibinfo {author} {\bibfnamefont {T.}~\bibnamefont
  {Pohl}}\ and\ \bibinfo {author} {\bibfnamefont {P.~R.}\ \bibnamefont
  {Berman}},\ }\bibfield  {title} {\enquote {\bibinfo {title} {{B}reaking the
  {D}ipole {B}lockade: {N}early {R}esonant {D}ipole {I}nteractions in
  {F}ew-{A}tom {S}ystems},}\ }\href {\doibase 10.1103/PhysRevLett.102.013004}
  {\bibfield  {journal} {\bibinfo  {journal} {Phys. Rev. Lett.}\ }\textbf
  {\bibinfo {volume} {102}},\ \bibinfo {pages} {013004} (\bibinfo {year}
  {2009})}\BibitemShut {NoStop}%
\bibitem [{\citenamefont {Gurian}\ \emph {et~al.}(2012)\citenamefont {Gurian},
  \citenamefont {Cheinet}, \citenamefont {Huillery}, \citenamefont {Fioretti},
  \citenamefont {Zhao}, \citenamefont {Gould}, \citenamefont {Comparat},\ and\
  \citenamefont {Pillet}}]{Gurian2012}%
  \BibitemOpen
  \bibfield  {author} {\bibinfo {author} {\bibfnamefont {J.~H.}\ \bibnamefont
  {Gurian}}, \bibinfo {author} {\bibfnamefont {P.}~\bibnamefont {Cheinet}},
  \bibinfo {author} {\bibfnamefont {P.}~\bibnamefont {Huillery}}, \bibinfo
  {author} {\bibfnamefont {A.}~\bibnamefont {Fioretti}}, \bibinfo {author}
  {\bibfnamefont {J.}~\bibnamefont {Zhao}}, \bibinfo {author} {\bibfnamefont
  {P.~L.}\ \bibnamefont {Gould}}, \bibinfo {author} {\bibfnamefont
  {D.}~\bibnamefont {Comparat}}, \ and\ \bibinfo {author} {\bibfnamefont
  {P.}~\bibnamefont {Pillet}},\ }\bibfield  {title} {\enquote {\bibinfo {title}
  {Observation of a resonant four-body interaction in cold cesium rydberg
  atoms},}\ }\href {\doibase 10.1103/PhysRevLett.108.023005} {\bibfield
  {journal} {\bibinfo  {journal} {Phys. Rev. Lett.}\ }\textbf {\bibinfo
  {volume} {108}},\ \bibinfo {pages} {023005} (\bibinfo {year}
  {2012})}\BibitemShut {NoStop}%
\bibitem [{\citenamefont {C{\^o}t{\'e}}\ \emph {et~al.}(2006)\citenamefont
  {C{\^o}t{\'e}}, \citenamefont {Russell}, \citenamefont {Eyler},\ and\
  \citenamefont {Gould}}]{Cote2006}%
  \BibitemOpen
  \bibfield  {author} {\bibinfo {author} {\bibfnamefont {R.}~\bibnamefont
  {C{\^o}t{\'e}}}, \bibinfo {author} {\bibfnamefont {A.}~\bibnamefont
  {Russell}}, \bibinfo {author} {\bibfnamefont {E.~E.}\ \bibnamefont {Eyler}},
  \ and\ \bibinfo {author} {\bibfnamefont {P.~L.}\ \bibnamefont {Gould}},\
  }\bibfield  {title} {\enquote {\bibinfo {title} {{Q}uantum random walk with
  {R}ydberg atoms in an optical lattice},}\ }\href {http://stacks.iop.org/1367-2630/8/i=8/a=156} {\bibfield  {journal} {\bibinfo
   {journal} {New J. Phys.}\ }\textbf {\bibinfo {volume} {8}},\ \bibinfo
  {pages} {156} (\bibinfo {year} {2006})}\BibitemShut {NoStop}%
\bibitem [{\citenamefont {W\"uster}\ \emph {et~al.}(2010)\citenamefont
  {W\"uster}, \citenamefont {Ates}, \citenamefont {Eisfeld},\ and\
  \citenamefont {Rost}}]{Wuster2010}%
  \BibitemOpen
  \bibfield  {author} {\bibinfo {author} {\bibfnamefont {S.}~\bibnamefont
  {W\"uster}}, \bibinfo {author} {\bibfnamefont {C.}~\bibnamefont {Ates}},
  \bibinfo {author} {\bibfnamefont {A.}~\bibnamefont {Eisfeld}}, \ and\
  \bibinfo {author} {\bibfnamefont {J.~M.}\ \bibnamefont {Rost}},\ }\bibfield
  {title} {\enquote {\bibinfo {title} {{N}ewton's {C}radle and {E}ntanglement
  {T}ransport in a {F}lexible {R}ydberg {C}hain},}\ }\href {\doibase 10.1103/PhysRevLett.105.053004} {\bibfield  {journal} {\bibinfo  {journal}
  {Phys. Rev. Lett.}\ }\textbf {\bibinfo {volume} {105}},\ \bibinfo {pages}
  {053004} (\bibinfo {year} {2010})}\BibitemShut {NoStop}%
\end{thebibliography}

%

\end{document}